\documentclass[sigconf]{acmart}
\usepackage{fancyhdr}
\usepackage{enumitem}
\usepackage{bm}

\usepackage{amsmath,amsfonts,bm}

\def\eqref#1{equation~\ref{#1}}

\def\1{\bm{1}}

\def\ve{{\bm{e}}}

\def\vz{{\bm{z}}}

\def\mE{{\bm{E}}}

\def\mZ{{\bm{Z}}}

\DeclareMathAlphabet{\mathsfit}{\encodingdefault}{\sfdefault}{m}{sl}
\SetMathAlphabet{\mathsfit}{bold}{\encodingdefault}{\sfdefault}{bx}{n}

\usepackage{multirow}
\usepackage{subfigure}
\usepackage{amsmath}  %
\usepackage{cases}  %
\usepackage{balance}
\usepackage{wrapfig}
\usepackage[titletoc]{appendix}
\usepackage[]{fixltx2e}
\usepackage[bottom]{footmisc} %
\AtBeginDocument{%
  \providecommand\BibTeX{{%
    \normalfont B\kern-0.5em{\scshape i\kern-0.25em b}\kern-0.8em\TeX}}}

\newcommand{\revise}[1]{\textcolor{black}{#1}}

\author{Yunzhu~Pan$^{*}$}
\affiliation{%
  \institution{
  University of Electronic Science and Technology of China}
  \city{Chengdu}
  \country{China}}
  
\author{Chen~Gao$^{\dagger}$}
\affiliation{%
  \institution{
  Beijing National Research Center for Information Science and Technology,\\ Department of Electronic Engineering, Tsinghua University}
  \city{Beijing}
  \country{China}}

\author{Jianxin~Chang}
\affiliation{%
  \institution{Beijing Kuaishou Technology Co., Ltd.}
  \city{Beijing}
  \country{China}}

\author{Yanan~Niu}
\affiliation{%
  \institution{Beijing Kuaishou Technology Co., Ltd.}
  \city{Beijing}
  \country{China}}
  
\author{Yang~Song}
\affiliation{%
  \institution{Beijing Kuaishou Technology Co., Ltd.}
  \city{Beijing}
  \country{China}}
  
\author{Kun~Gai}
\affiliation{%
  \institution{Unaffiliated}
  \city{Beijing}
  \country{China}}

\author{Depeng~Jin}
\affiliation{%
  \institution{
  Beijing National Research Center for Information Science and Technology,\\ Department of Electronic Engineering, Tsinghua University}
  \city{Beijing}
  \country{China}}
  
\author{Yong~Li}
\affiliation{%
  \institution{
  Beijing National Research Center for Information Science and Technology,\\ Department of Electronic Engineering, Tsinghua University}
  \city{Beijing}
  \country{China}}

\thanks{$*$Work done when interning at Tsinghua University.\\
$\dagger$Chen Gao is the corresponding author (chgao96@gmail.com).}

\renewcommand{\authors}{Yunzhu Pan, Chen Gao, Jianxin Chang, Yanan Niu, Yang Song, Kun Gai, Depeng Jin, Yong Li}

\renewcommand{\shortauthors}{Pan, et al.}

\copyrightyear{2023}
\acmYear{2023}
\setcopyright{acmlicensed}\acmConference[RecSys '23]{Seventeenth ACM
Conference on Recommender Systems}{September 18--22, 2023}{Singapore,
Singapore}
\acmBooktitle{Seventeenth ACM Conference on Recommender Systems (RecSys
'23), September 18--22, 2023, Singapore, Singapore}
\acmPrice{15.00}
\acmDOI{10.1145/3604915.3608814}
\acmISBN{979-8-4007-0241-9/23/09}

\begin{document}

\title{Understanding and Modeling Passive-Negative Feedback for Short-video Sequential Recommendation}

\begin{abstract}
Sequential recommendation is one of the most important tasks in recommender systems, which aims to recommend the next interacted item with historical behaviors as input.
Traditional sequential recommendation always mainly considers the collected positive feedback such as click, purchase, etc.
However, in short-video platforms such as TikTok, video viewing behavior may not always represent positive feedback. Specifically, the videos are played automatically, and users passively receive the recommended videos.
In this new scenario, users passively express negative feedback by skipping over videos they do not like, which provides valuable information about their preferences.
Different from the negative feedback studied in traditional recommender systems, this passive-negative feedback can reflect users' interests and serve as an important supervision signal in extracting users' preferences.
Therefore, it is essential to carefully design and utilize it in this novel recommendation scenario.
In this work, we first conduct analyses based on a large-scale real-world short-video behavior dataset and illustrate the significance of leveraging passive feedback. 
We then propose a novel method that deploys the sub-interest encoder, which incorporates positive feedback and passive-negative feedback as supervision signals to learn the user's current active sub-interest. 
Moreover, we introduce an adaptive fusion layer to integrate various sub-interests effectively.
To enhance the robustness of our model, we then introduce a multi-task learning module to simultaneously optimize two kinds of feedback -- passive-negative feedback and traditional randomly-sampled negative feedback.
The experiments on two large-scale datasets verify that the proposed method can significantly outperform state-of-the-art approaches.
The code is released at \textit{\url{https://github.com/tsinghua-fib-lab/RecSys2023-SINE}}.

\end{abstract}



\maketitle

\section{Introduction}\label{sec::intro}

Sequential recommendation is one of the most fundamental tasks, which enhances the basic collaborative filtering with the sequential behaviors~\cite{wu2022survey}.
In real-world information systems, sequential recommenders serve as the core of the recommendation engine.
The traditional methods of sequential recommendation follow the same paradigm of learning user interests from the behavioral sequence, with positive feedback only, such as~\cite{kang2018self, yu2019adaptive, hidasi2015session}.
However, the recent success of short-form videos such as TikTok\footnote{https://www.tiktok.com/en/} has re-defined the interaction manner of how the user access online content.
\revise{
In these platforms, users passively receive recommended items in a new single-column layout, and the videos are automatically played unless users actively skip over them. 
As a result, a new type of passive-negative feedback has emerged, where users passively skip videos to find content that interests them. 
However, this passive-negative feedback cannot be handled in the traditional manner of negative feedback in the previous works~\cite{wang2022unbiased, gong2022positive, seo2022siren, DBLP:conf/ijcai/XieLWWXL20}.
Specifically, watching a video on these platforms may not necessarily indicate the user likes the video, as videos are played automatically, but users can choose to skip over it, which suggests that the user is not interested in this particular video.
Therefore, there is a need for novel approaches to handle this unique feedback scenario.
}

\revise{We first conduct empirical analysis from a statistical perspective and the recommendation performance to fully understand this new negative feedback.}
Our results are based on the collected behavioral dataset from one of the largest short-video platforms, along with the state-of-the-art sequential recommendation model.
The empirical analysis\footnote{The experiments details are presented in Section~\ref{sec::motivation}.} first demonstrates that the negative feedback in real-world applications is always passive, \textit{i.e.,} the users choose to skip over the recommended content, and often do not explicitly report it to the platform.
Second, the analysis results show that roughly using these kinds of passive-negative feedback in the recommendation model will even lead to a significant performance drop, illustrating a challenge of feedback learning.
Last, based on the hierarchical categorical information, we find that the items receiving passive-negative feedback \revise{often share similar categories with} those that receive positive feedback. That is, the items are not truly hated/disliked by users. On the contrary, users are not satisfied with \revise{certain features} of the items. \revise{For example}, users may have consumed similar content before and no longer find it useful or interesting. \revise{Therefore, without a unique design, it is difficult for a model to learn the difference between passive-negative and positive items.}

The above results show the significance of negative feedback and illustrate the challenge and its explanation of taking this passive-negative feedback into consideration of the sequential recommenders.
To address it, we propose a novel method named SINE (short for \textbf{S}ub-\textbf{IN}terest learning with Negativ\textbf{E} feedback).
Specifically, we first design a mix-feedback sequential encoder that takes both positive feedback and passive-negative feedback, and extracts sub-interests for the given context with a sub-interest-based self-attention layer.
We then propose an adaptive fusion layer that selects the activated sub-interests and deactivates the remaining ones in the user behaviors.
Last, to well exploit both the collected passive-negative feedback and the unobserved ones, we adopt a multi-task learning paradigm for the optimization of model parameters.
The main contributions of this paper can be summarized as follows,
\begin{itemize}[leftmargin=*]
	\item 
 \revise{In this paper, we approach the new problem of}
 understanding and modeling of user's passive-negative feedback in the sequential recommendation\revise{, particularly in the new paradigm of single-column short-video platforms}. \revise{We collect users' skipping behaviors from short-video platforms as passive-negative feedback, which provide a more accurate reflection of users' preferences. By skipping a video, users actively indicate their disinterest or dissatisfaction with the content, which provides a clear signal to the recommender system about their preferences. This approach contrasts with previous studies that have relied on exposure-based negative feedback, such as "expose-but-not-click" behavior, which may not always accurately reflect a user's preferences due to the exposure bias problem.}
 We conduct an empirical analysis on a large-scale real-world dataset, which reveals the importance, challenge, and explanations for negative feedback in today's short-video platforms.
	\item We proposed a method with sub-interest learning, which can well handle the mix-feedback sequence and extract the sub-interests that belong to different subspaces. The joint optimization elegantly takes positive feedback, negative feedback, and unobserved feedback into consideration at the same time.
	\item We conduct experiments on two large-scale datasets from two mainstream short-video platforms. Extensive results show that our method can steadily and significantly outperform the state-of-the-art recommendation methods.
	Our further experiments well demonstrate the rationality of each component of our SINE method. The results of recommendation performance also correspond well to our earlier data analysis.
\end{itemize}

The remainder of this paper is as follows.
We first conduct an empirical analysis on the real-world dataset and provide the motivations of the research problem in Section~\ref{sec::motivation}.
We then formally define the research problem in Section~\ref{sec::probdef} and present our solution in Section~\ref{sec::method}.
We conduct experiments on Section~\ref{sec::experiments} and review the related works in Section~\ref{sec::related}.
Last, we conclude this paper and discuss the important future works in Section~\ref{sec::conclusion}.

\section{Data Analysis and Motivation}\label{sec::motivation}

In this section, we aim to understand the significance of negative feedback and the critical challenge of modeling it in real-world short-video recommendation.
Specifically, we conduct analysis on the behavioral dataset collected from one of the largest micro-video platforms.
We first present the ratio of negative behavior compared with other kinds of behaviors, through which we can find the active-negative behavior is very sparse, and the passive-negative behaviors are far easier to collect.
We then conduct experiments on the widely-used sequential recommendation model to illustrate the challenge of leveraging the passive-negative feedback data.
Finally, we further study the passive-negative feedback to answer how this behavior occurs, serving as guidance for designing more powerful recommendation models.

\subsection{Data characteristics of negative feedback}

\begin{figure}[t!]
    \centering
    \subfigure[User feedback distribution.]{
    \includegraphics[width=0.45\linewidth]{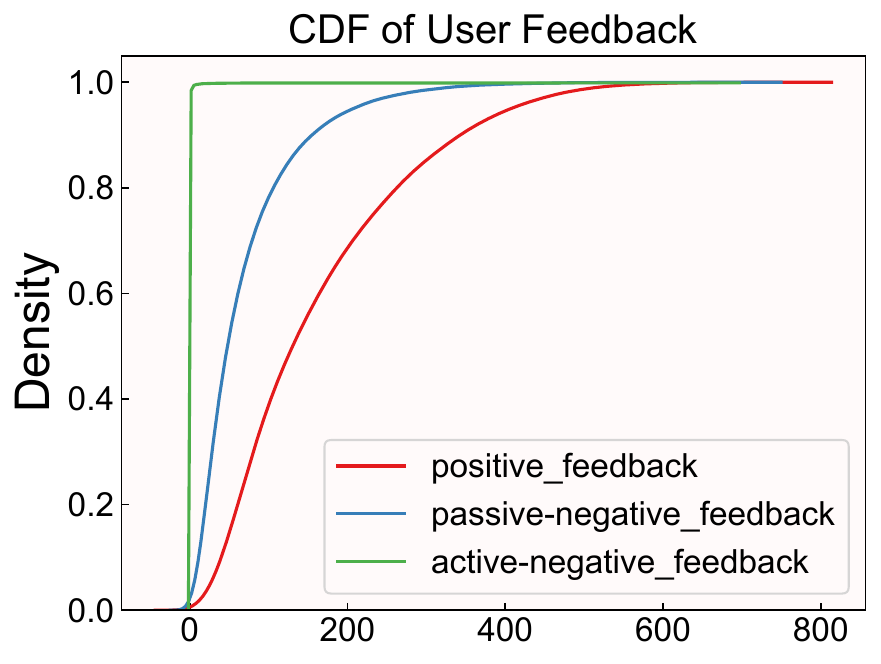}
    }
    \subfigure[Item feedback distribution.]{
    \includegraphics[width=0.47\linewidth]{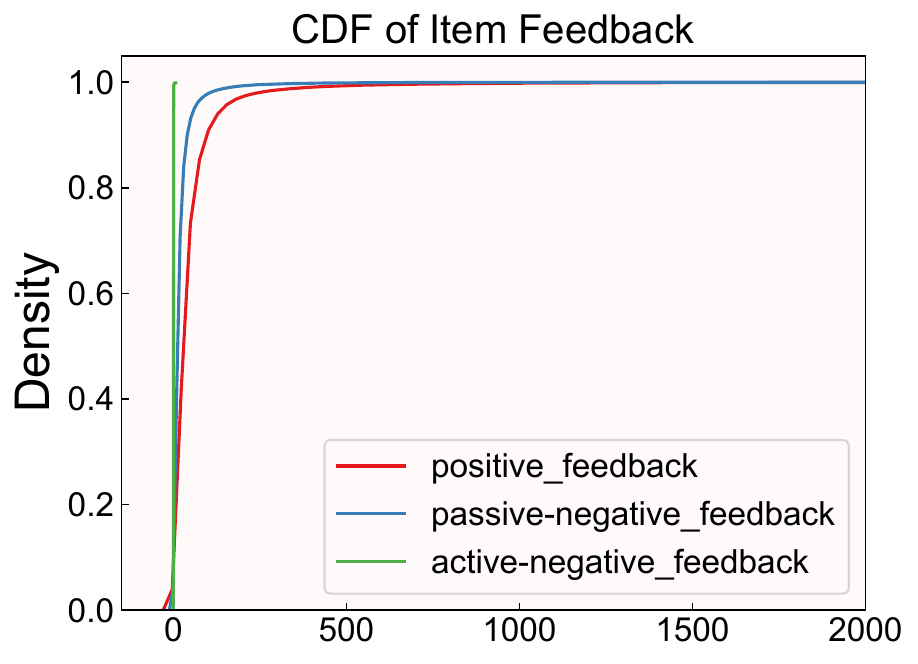}
    }
    \caption{Distribution of three kinds of feedback on the Kuaishou dataset.}
    \label{fig:category_compare_kuaishou}
\end{figure}

We first obtain the characteristics of negative feedback, which consists of two major forms in a typical recommendation in today's information system.
First, the user can actively convey their opinions by selecting options including \textit{``not interested''}, \textit{``reduce similar recommendations''}, or even \textit{``hate''}.
Second, when the user passively receives sequentially recommended videos, the user can skip over the video fast if he/she feels not interested.
For these two kinds of negative feedback, we present the interaction number based on a real-world dataset collected from one of the famous short-video platform\footnote{This dataset will be used for evaluation dataset, and its detail will be introduced in Section~\ref{sec::experiments}.}, along with effective-view behavior, \textit{i.e.}, the user has spent adequate time in watching the video\footnote{A common criterion in the industry is the half length of the whole video.} in Fig.~\ref{fig:category_compare_kuaishou}.
\revise{From the figure, we can observe that the active-negative feedback is far sparser than the passive-negative feedback, and the density of passive-negative feedback is close to that of positive behavior.}
The extremely sparse active-negative feedback can be easily understood that users are not always willing to perform additional operations.
It means that when designing recommendation models, active-negative feedback can hardly enhance the preference modeling as it is too sparse, while passive-negative feedback may play an important role.

\subsection{Study of leveraging passive-negative feedback}

\begin{figure*}[t!]
	\centering
	\subfigure[Results of introducing passive-negative feedback on Kuaishou.]{
		\includegraphics[width=0.30\linewidth]{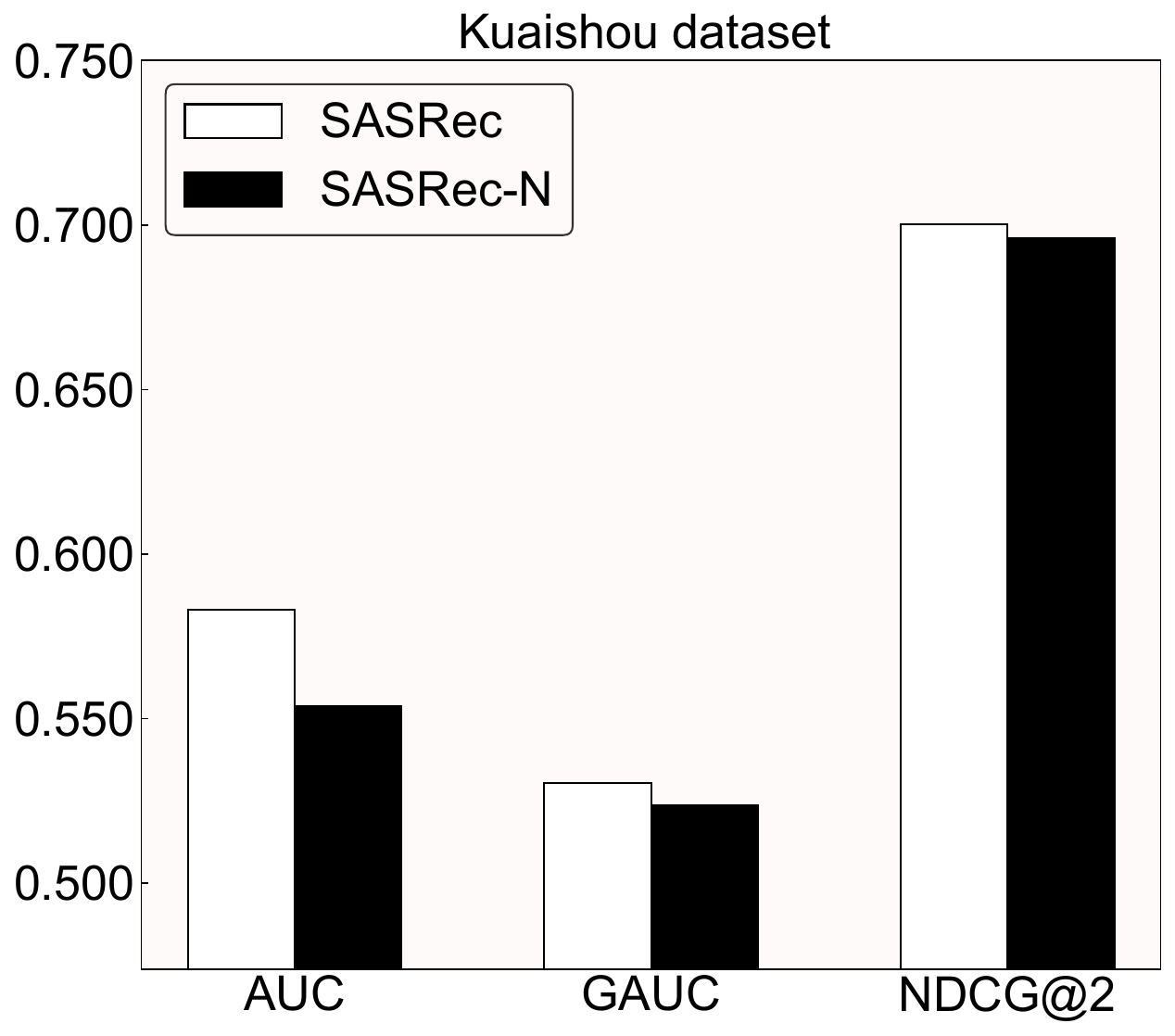}
	}
	\subfigure[Results of introducing passive-negative feedback on WeChat.]{
		\includegraphics[width=0.30\linewidth]{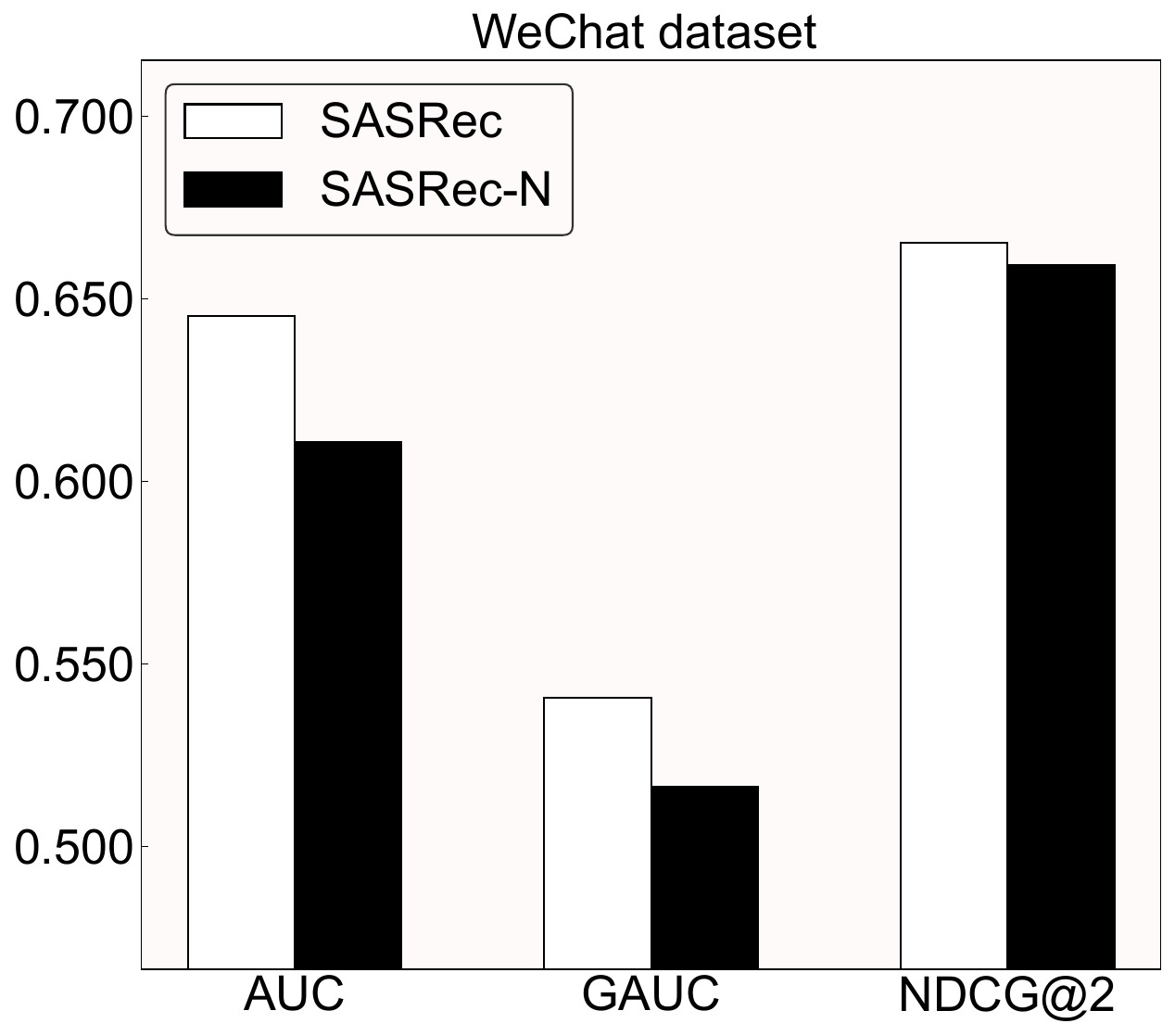}
	}
	\subfigure[Occurrence of passive-negative feedback (case 1: different level-1 categories; case 2: same level-1 but different level-2; case 3: same level-2 but different level-3; case 4: same level-3.)]{
	\includegraphics[width=0.30\linewidth]{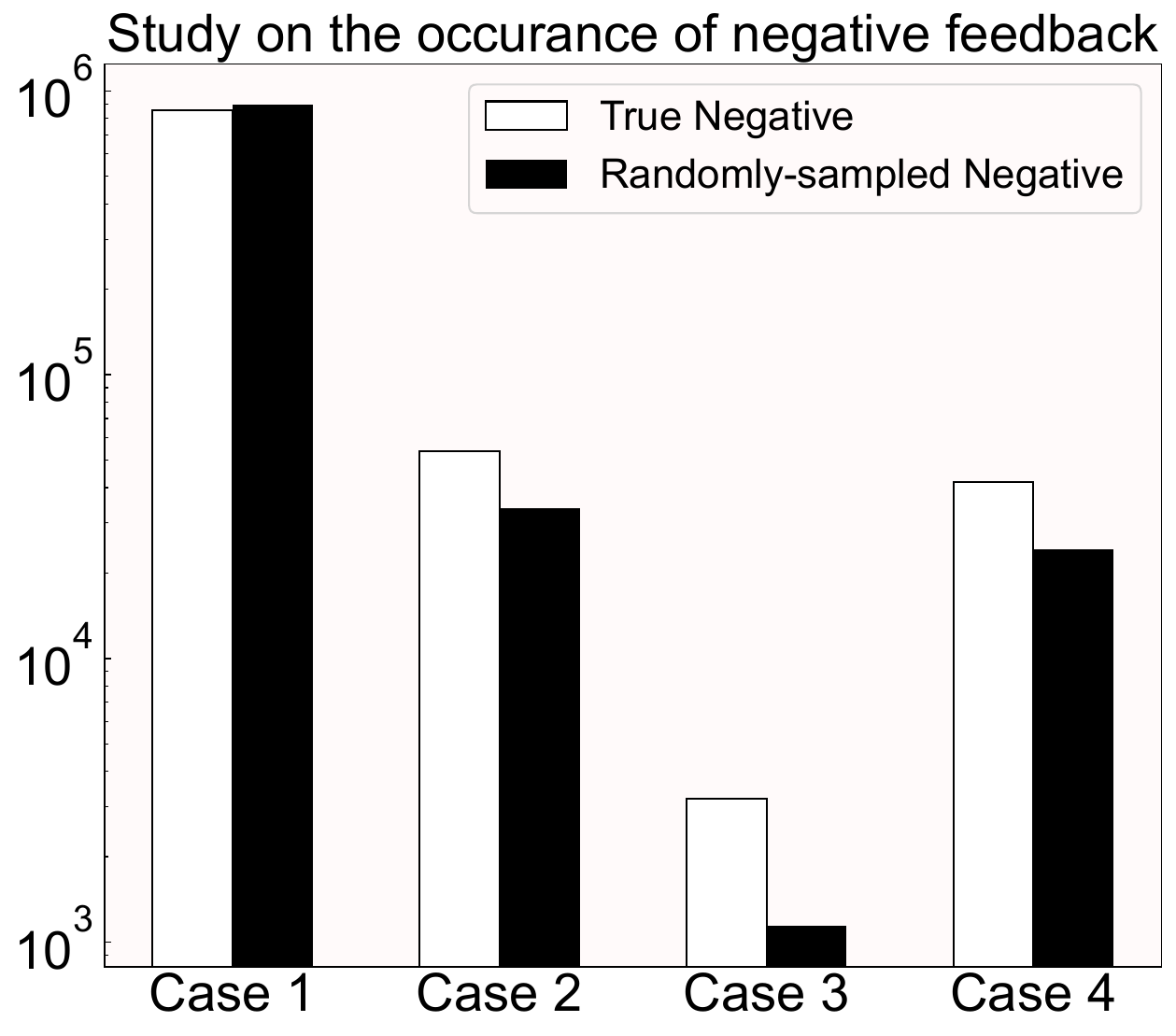}
}
	\caption{Data analysis of the passive-negative feedback on real-world datasets via comparing recommendation models (a)(b) and the category-based statistics (c).}
	\label{fig:XXX}
\end{figure*}

However, despite the promising density, the passive-negative feedback is not easy to handle in recommender systems. In this section, we want to design an experiment to demonstrate the challenge of leveraging such kind of feedback.
From the perspective of representation learning, the recommendation models need to project users and items into the latent vector in low-dimensional space, and the matching scores between vectors can be used for generating recommendation results.
The optimization of the representation learning can treat the passive-negative feedback as a negative sample, no matter with point-wise optimization~\cite{he2016fast,xue2017deep} or pair-wise optmization~\cite{rendle2012bpr}.
Therefore, we conduct experiments with the popular SASRec~\cite{kang2018self} model as the backbone, which adopts self-attention layers to encode users' behavioral sequences.
Specifically, in the first case, we use the observed view behavior as positive feedback and randomly sampled items (without observed behavior) as negative feedback.
That is, the first case refers to the normal learning procedure.
In the second space, we also consider the truly-observed passive-negative behavior as the negative sample, and we carefully control its ratio compared with randomly-sampled samples by a hyper-parameter.
\revise{We name it SASRec-N (N denotes Negative).}
We then obtain the performance of two cases, both of which are under the careful and extensive hyper-parameter search.
The results are shown in Fig.~\ref{fig:XXX}, in which three widely-considered metrics, AUC, GAUC, and NDCG, are used.
We can observe from the results that the intuitive design of leveraging passive-negative feedback in the negative sampling and optimization procedure can even harm the recommendation performance.
The results demonstrate that it is challenging to well exploit passive-negative feedback even if it seems to make sense.

\subsection{Analysis of how passive-negative feedback occurs}

Given the analysis above, which shows the importance along with the challenges of exploiting the passive-negative feedback, 
\revise{we further explore the reasons behind the occurrence of passive-negative feedback to better address this challenge and inspire model design.}
We try to answer the question of why the user decides to skip over the video, which is very hard as we can hardly know what the users are thinking about.
To address it, we choose to answer a similar but easier question: what is the difference between positive feedback and negative feedback if they occur together?
Therefore, we introduce the auxiliary category information of videos and analyze the difference between positive-negative feedback on the category aspect.
\revise{Specifically, we analyze the categorical difference between positive feedback and two kinds of negative feedback - true negative feedback (\textit{i.e.}, passive-negative feedback) and randomly-sampled negative feedback, based on three levels of categories from coarse-grained to fine-grained. The categorical differences from real-world interaction data are shown in Fig.~\ref{fig:XXX}(c).}
\revise{To better compare the differences in categories at different granularities, we divide the results into four cases, explained as follows.}
Case 1 denotes the negative feedback video having a different level-1 category from the positive feedback video. Case 2 means the positive feedback video and negative feedback video have the same level-1 category but differ in the category level-2. Case 3 indicates that the positive feedback video and negative feedback video have the same level-1 and level-2 category but different level-3 category. Case 4 shows the number of positive and negative feedback videos sharing the same category at all three levels.
We can observe from the results that the categories of a positive-negative pair (occurs very close) are always very similar or even the same.
\revise{In addition, the number of positive feedback and true  negative feedback having the same category is larger than that of positive feedback and randomly-sampled negative feedback.}
\revise{This difference is more evident in finer-grained categories.}
\revise{Therefore, we can infer that the recommender system has well estimated the user's interests as the recommended item that received negative feedback is already very similar to the positive item. However, there may be some sub-aspects of the item that do not align with the user's preferences, resulting in the final negative behavior. In other words, the passive-negative item has met a part of the user's preferences but fails to match all of them, causing the user to skip over the item.}

\vspace{0.2cm}

In short, we conduct early analysis on a real-world dataset, which first shows the importance of passive-negative feedback in preference learning, then illustrates the challenges of exploitation via experimental results, and finally partly provides the reasons for passive-negative feedback.
\section{Problem Formulation}\label{sec::probdef}

\begin{table*}[t] \label{tab:notations}
    \centering
    \caption{Frequently used notations in this paper.}
    \label{tab:notations}
    \begin{tabular}{ll}\toprule
         Notations & Descriptions\\\hline
         $|\cdot|$ & The cardinality of a set\\
         $\langle \cdot \rangle$ & Inner product\\
         $\mathcal{U}, \mathcal{I}$ & The set of users and items \\
         $M$, $N$ & The number of users and items \\
         $\mathbf{S}^u$, $\mathbf{R}^u$ & The item sequence and the feedback label of user $u$ \\ 
         $\mathbf{S}^u_+$, $\mathbf{S}^u_-$ & The item sequence with positive/passive-negative feedback of user $u$\\ 
         $\mE \in \mathbb{R}^{N \times D}$ & Item embedding matrix with dimension size of $D$ \\
         $\ve^u_+, \ve^u_-$ &  The item embedding of positive/passive-negative feedback of user $u$\\ 
         $\mZ$ & Sub-interest prototypes\\
         $K$ & Number of sub-interest prototypes\\
        $\hat{z}^{s^{u}_{+, 1}}, \cdots, \hat{z}^{s^{u}_{+, |S^u_+|}}$ & The corresponded sub-interest of the positive feedback sequence of user $u$\\
        $\mathbf{Q}, \mathbf{K}, \mathbf{V}, \mathbf{P}$ & Query/ Key/Value/Position embedding in self-attention\\
        $\alpha$ & Self-attention weight\\
        $\beta$ & Sub-interests weight\\
        $\widetilde{\textbf{o}}_t^{u, k}$ & The $k$-th sub-interest of user $u$ at timestamp $t$\\
        $r, \gamma$ & weight and normalized weight of different sub-interests\\
        $L_1, L_2$  & Loss function on two kind of pairs $O_1$ and $O_2$\\
        $L_{dis}$ & The discrepancy loss on sub-interest prototypes\\
         \bottomrule
    \end{tabular}
\end{table*}

\revise{In this work, we approach the new problem of modeling the passive-negative feedback and positive feedback in the short-video sequential recommendation.}
Based on the analysis of the real-world data in Section~\ref{sec::motivation}, the passive-negative feedback is more promising than the active-negative, considering the data sparsity.
Let $\mathcal{U}$ and $\mathcal{I}$ denote the sets of users and items, respectively, of which the sizes are $M$ and $N$.
Given a user $u \in \mathcal{U}$, its sequence can be denoted as $\mathbf{S}^u = \{s^u_1, s^u_2, \cdots,  s^u_{|\mathbf{S}^u|}\}$.
Since in the behavioral sequence, there are two kinds of feedback, we introduce another sequence $\mathcal{R}^u = \{r^u_1, r^u_2, \cdots,  r^u_{|\mathbf{R}^u|}\}$, of which $r_u$ could be $1$ for positive feedback and $0$ for passive-negative feedback. The frequently-used symbols are listed in Table~\ref{tab:notations}.
Therefore, the new-form sequential recommendation can be formally defined as follows.

\noindent \textbf{Input:} The sequentially-interacted items of the user $u$,  $\mathbf{S}^u$, along with the feedback type,  $\mathcal{R}^u$.

\noindent \textbf{Output:} A recommendation model that can estimate the probability, $p^u_{|\mathbf{S}^u|+1, i}$ , that the given user $u$ will interact with the target item $i$ at the next time.

\begin{figure*}
    \centering
    \includegraphics[width=0.85\linewidth] {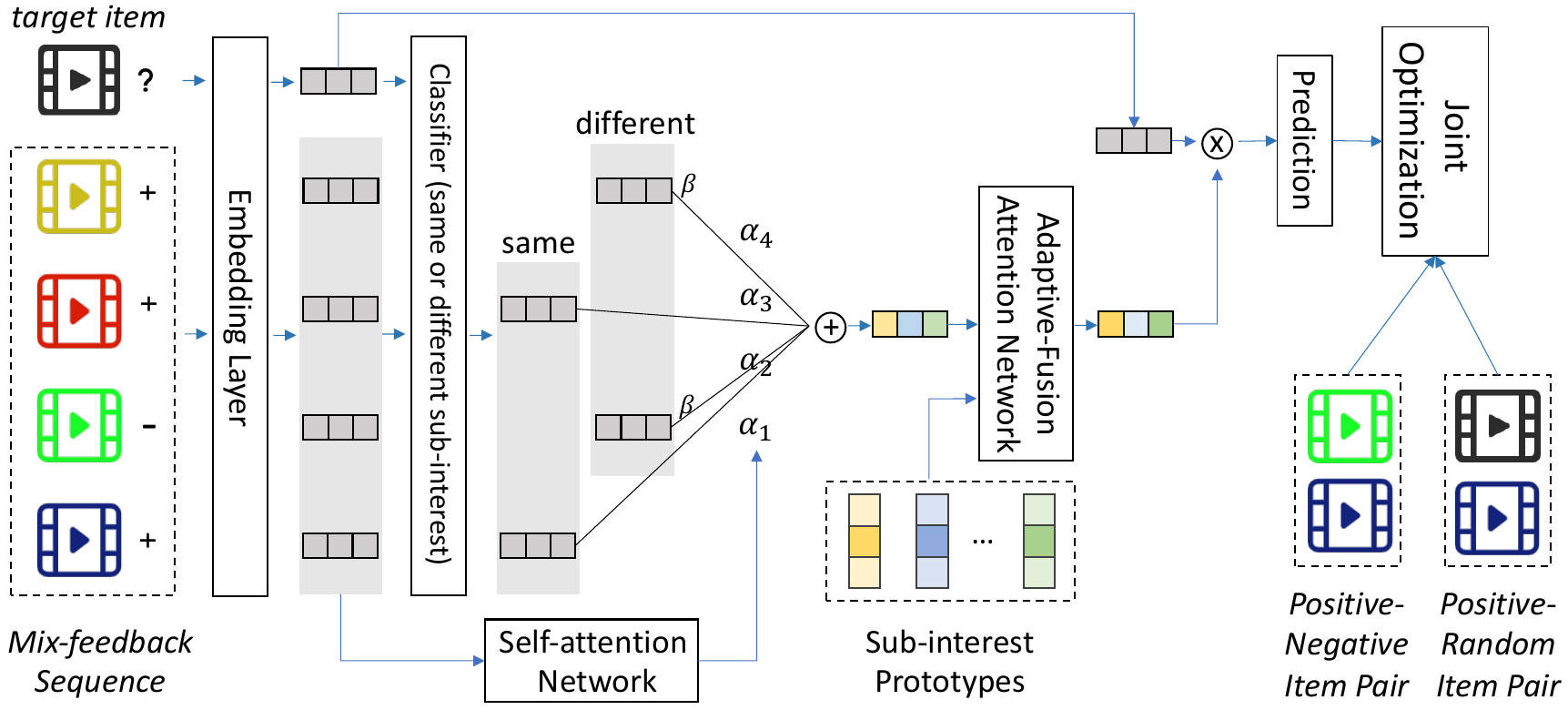}
    \caption{Illustration of our proposed SINE method.}
    \label{fig:framework}
\end{figure*}

\section{Methodology}\label{sec::method}
Inspired and motivated by the analysis of the negative feedback, we propose our method named SINE based on the idea that the passive-negative feedback is caused by the mismatch of specific sub-interests. 
\revise{
Specifically, as mentioned above, the videos that received negative feedback share similar categories with the positive videos, which indicates the recommender system has only partially met the user's preferences but fails to match all of them, causing users to skip over.
}
Our approach, as illustrated in Fig.~\ref{fig:framework}, consists of the following components:
\begin{itemize}[leftmargin=*]
    \item \textbf{Sub-interest-based sequential encoder.} Different from the pure-positive feedback in traditional sequential recommenders, we propose to encode the complex sequence with mixed feedback with a self-attention layer.
    We propose to project user preferences into multiple sub-spaces, each corresponding to a specific aspect that may lead to user behaviors.
    \item \textbf{Adaptive-fusion prediction layer.} 
    With the learned sub-interests, we deploy a prediction layer that can distinguish not only the normal negative feedback but also the passive-negative feedback from positive feedback via learned weights-based interest fusion.
    \item \textbf{Joint optimization.}
    To well leverage these two kinds of feedback, we propose to jointly optimize the model parameters under a multi-task learning paradigm, in which fitting each kind of data can be treated as a task, and the hyper-parameters can well control the importance of two tasks.
\end{itemize}

\subsection{Sub-interest-based Sequential Encoder}
\subsubsection{Embeddings of users, items, and sub-interest prototypes.}
First, we build an embedding matrix $\mE \in \mathbb{R}^{N \times D}$ \revise{($D$ is the dimension size)} that assigns low-dimensional vectors to encode each item and then retrieved embedding of the user sequence $S^u$ at the timestamp $t$ can be represented as follows,
\begin{equation}
\mE^u = (\ve^u_1, \ve^u_2,  \cdots, \ve^u_{t}),
\end{equation}
Since the user sequence $S^u$ contains two kinds of feedback, and, for convenience, we use two symbols $S_+^u$ and  $S_-^u$ to denote the positive feedback sequence and passive-negative feedback sequence, respectively, defined as follows,
\begin{equation}
\begin{aligned}
S_+^u \in S^u, \text{where~}  R^u=1, \\
S_-^u \in S^u, \text{where~}  R^u=0,
\end{aligned}
\end{equation}
and item embedding of positive feedback and passive-negative feedback is denoted as $\ve^u_+$ and $\ve^u_-$, respectively.

Generally, a user interacts with a short video due to multiple aspects, such as the style, author, music, etc.
Thus, the users can refuse the recommendation by skipping over the video only because one of these aspects does not satisfy the user.
To capture the sub-interests towards different aspects, we build the prototype vectors, each of which represents an aspect of users' interests, shown as follows,
\begin{equation} 
\mZ = (\vz_1, \vz_2, \cdots, \vz_K),
\end{equation}
where $K$ denotes the number of prototypes, which is a controllable hyper-parameter.

To map each feedback in a positive feedback sequence to the corresponding sub-interest (\textit{i.e.,} find the dominate sub-interest that leads to the behavior), we propose to utilize the recent passive-negative feedback to calculate the gap between the matching scores of positive and passive-negative feedback with each sub-interest, calculated as follows,
\revise{
\begin{equation}
\hat{z}^i =  \mathop{\arg\max}\limits_{k} \ve^u_+ \cdot \vz_k - \ve^u_- \cdot \vz_k,
\end{equation}
}
\revise{
where $\hat{z}^i$ is selected from $1, 2, \cdots, K$ sub-interests. Then the corresponding sub-interests of positive feedback in a user sequence are as follows, 
\begin{equation}
\hat{z}^{s^{u}_{+, 1}}, \hat{z}^{s^{u}_{+, 2}}, \cdots, \hat{z}^{s^{u}_{+, |S^u_+|}},
\end{equation}
}
For positive feedback without recent negative feedback, we omit the second term and directly calculate the similarity between item and prototype embeddings. 

\subsubsection{Sub-interest enhanced sequence encoder.}
With the proposed component above, we obtain the specific aspect of user preferences that lead to user feedback, \textit{i.e.,} sub-interest, which can be further exploited to encode the behavioral sequence.
Specifically, we propose a multi-head self-attention-based encoder that can well distinguish the complex relations between items, especially for the impact of two kinds of feedback and 
the sub-interests.
The encoder can be formulated as follows,
\begin{equation}
\begin{aligned}
\mathbf{X}=\mathbf{E}^u+\mathbf{P}, \\
\mathbf{Q}=\mathbf{X}\mathbf{W^q}; \mathbf{K}=\mathbf{X}\mathbf{W^k}; \mathbf{V}=\mathbf{X}\mathbf{W^v},
\end{aligned}
\end{equation} 
where $\mathbf{P}$ denotes the position embedding, $\mathbf{W^q}, \mathbf{W^k}, \mathbf{W^v} \in \mathbb{R}^{D \times D}$ are three projection matrices and $\mathbf{Q}, \mathbf{K}, \mathbf{V} \in \mathbb{R}^{L \times D}$ ($L$ is the length of sequence).

For the sub-interest space, there exist implicit relations between different sub-interests, and thus we design a controllable weight $\beta$ that models the correlation of sub-interests.
\revise{First, the self-attention weights for each item are formulated as follows:}
\begin{equation}
	\alpha = \textbf{Softmax}(\frac{\mathbf{Q} \mathbf{K}^T}{\sqrt{d}}),
\end{equation}
\revise{Then, we introduce $\beta$ to learn the correlation of sub-interests, defined as follows:}
\begin{equation}
		\beta =\left\{ 
			\begin{aligned}
				\beta_1, \hat{z}^i=\hat{z}^{s^{u}_{+, |S^u_+|}}, \\
				\beta_2, \hat{z}^i \neq \hat{z}^{s^{u}_{+, |S^u_+|}},
	\end{aligned}
\right.
\end{equation}
where $\beta_1+\beta_2=1$.
\revise{Last, we obtain the final attention weights for each item by multiplying $\beta$ to $\alpha$, shown as below:}
\begin{equation}
    \hat{\alpha} = \beta \cdot \alpha,
\end{equation}
\revise{With the obtained $\hat{\alpha}$, we not only consider the items' role in the whole interaction sequence but also take into account the relationships between sub-interests.
Then, we can calculate the encoded vector as follows,}
\begin{equation}
    \textbf{o}^u_t = f_\textbf{FFN}(\textbf{Norm}(\hat{\alpha} \textbf{V} + \textbf{X})) + \textbf{Norm}(\hat{\alpha} \textbf{V} + \textbf{X}),
\end{equation}
where $f_\textbf{FFN}$ denotes  a fully-connected layer, and $\textbf{Norm}$ means the layer normalization.

\revise{After we obtain the encoded user embedding $\textbf{o}^u_t$, we further project it into the pre-defined sub-interest prototypes space to generate a user-specific sub-interests embeddings, which is based on the context of the user's interaction history. The formulation is as follows, }
\begin{equation}
	\widetilde{\textbf{o}}_t^{u, k} = \textbf{o}^{u}_{t} + \sigma (\textbf{o}^{u}_{t} * \vz_k) \cdot \textbf{o}_t^u, 
\end{equation}
\revise{where $k$ means the $k$-th sub-interest. After the projection, each user will have $K$ user-specific sub-interests, which is the same number as the sub-interest prototypes.}

\subsection{Fusion Prediction Layer}
As for the real-world information system, user behavior is triggered by multiple aspects, which we have learned the corresponding representations in sub-interest spaces. 
To accurately predict the next item that users will interact with, it is essential to determine which aspect will play a vital role in users' current state.
Therefore, we propose to first estimate the importance of sub-interests before fusing them.
Specifically, we propose an attention network-based approach as follows,
\begin{equation}
	r_k = \sigma(\textbf{W}[\mathbf{e}_i; \widetilde{\textbf{o}}_t^{u, k}] + \textbf{b}),
\end{equation}
where $\textbf{W}$ and $\textbf{b}$ are learnable parameters.
The weights are further normalized as follows:
\begin{equation}
	\gamma_k = \frac{r_k}{\sum_{k=1}^{K}{r_k}}.
\end{equation}
Then the prediction score based on the learned weights and sub-interests can be formulated as follows,
\begin{equation}
	\textbf{Score}(u, i, t) = \sum_{k=1}^{K}(\gamma_k \cdot \widetilde{\textbf{o}}_t^{u, k} \mathbf{e}_i),
\end{equation}
where $\mathbf{e}_i$ denotes the embedding of the target item.

\noindent \textbf{Discussion.}
The learnable weights above are actually the generalized version of an intuitive design that only considers one sub-interest for a given behavior.
For example, if the weights are $1$ for one sub-interest and $0$ for others, the specific sub-interest dominates the user behavior.

\subsection{Joint Optimization}
From the perspective of the user-algorithm feedback loop, the passive-negative feedback is collected based on the exposure of the already-deployed recommendation algorithms. 
Therefore, it is essential to leverage all kinds of behaviors to optimize the parameters.
In our problem, there is additional negative feedback, \textit{i.e.}, the truly-observed passive-negative feedback, besides the randomly sampled ones. Specifically, the traditional recommenders, no matter collaborative filtering or sequential recommendation, tend to sample unobserved items as negative ones for optimization.

Thus, we propose pairwise learning on both two pairs: \{\textit{positive feedback, randomly-sampled negative feedback}\} and 
\{\textit{positive feedback, truly-observed negative feedback}\}.

The pairwise loss function defined on the two kinds of pairs is as follows:
        \revise{
        \begin{equation}
		\begin{aligned}
		L_1 = \sum_{(u,i,j)\in O_1} -ln\sigma(y(u,i)-y(u,j)) ,\\
		L_2 = \sum_{(u,i,j)\in O_2} -ln\sigma(y(u,i)-y(u,j)) ,
		\end{aligned}
	\end{equation}}
where $O_1 = \{(u,i,j)|(u,i)\in R^+, (u,j)\in R^-\}$ denotes the set of training data, where $R^+$ represents observed  behavior and $R^-$ represents unobserved behavior; 
 $O_2 = \{(u,i,j)|(u,i)\in R^+, (u,j)\in R^-\}$ denotes the set of training data, where $R^+$ represents observed behavior and $R^-$ represents truly-observed passive-negative behavior. 
Here $\sigma(\cdot)$ denotes the sigmoid function.
We also should ensure the disentanglement across the sub-interest prototypes.
Therefore, we design another distance correlation loss~\cite{szekely2007measuring}, $L_{dis} = dCor(\mathbf{Z})$, which tries to maximize the distance between prototypes.

The final joint loss function is performed under a paradigm of multi-task learning~\cite{zhang2018overview}:
\begin{equation}
	L = \lambda_1 L_1 + \lambda_2 L_2 + \lambda_3 L_{dis},
\end{equation}
\revise{where $\lambda_1$, $\lambda_2$, and $\lambda_3$ are three hyper-parameters that can control the importance of each loss, and we have $\lambda_1 + \lambda_2 + \lambda_3 = 1$.}

\section{Experiments}\label{sec::experiments}
In this section, we conduct extensive experiments on two real-world datasets in order to answer the following three research questions (RQs).

\begin{table*}
	\caption{Statistics of the two datasets used in experiments.}
	\label{tab::dataset}
	\begin{tabular}{cccccccc}
		\toprule
		Dataset & Users & Items & Instances & Positive & Negative & Average Length \\ 
		\midrule
		WeChat & 19,901 & 94,910 & 6,200,308 & 4,050,193 & 2,150,114 & 311 \\
		Kuaishou & 37,497 & 126,293 & 9,049,176 & 6,399,423 & 2,649,753 & 241 \\
		\bottomrule
	\end{tabular}
\end{table*}

\begin{itemize}
	\item RQ1: Can our proposed SINE method outperform the state-of-the-art solutions for sequential recommendation on two real-world datasets?
	\item RQ2: How about the rationality of each component in our proposed SINE method? 
	\item RQ3: How do the introduced hyper-parameters affect the recommendation performance of our proposed SINE model?
\end{itemize}

\subsection{Experimental Settings}
\subsubsection{Datasets}
We conduct experiments on two large-scale datasets from two of the largest short-video platforms,  Kuaishou and WeChat. Data are collected from the feed under real-world scenarios.
Basic statistics of the two datasets are summarized in Table \ref{tab::dataset}, where we present the interaction number of both positive feedback and negative feedback. 
Please note that there is a minor difference in statistics compared with the data analysis in Section~\ref{sec::motivation} since, for model evaluation, we have conducted a data processing of the widely-used N-core filtering~\cite{yu2021self,cho2021learning}. N is set as 10 in our experiments.

\begin{itemize}[leftmargin=*]
	\item \textbf{WeChat}\footnote{\url{https://algo.weixin.qq.com/}}. 
	This dataset was released by the Big Data Challenge in the year 2021, hosted by the recommendation team of WeChat, and it contains the behavioral logs on WeChat Channels (short-video services in the WeChat app) with a time period of two weeks.
We select the last and second-last interactions as validation and test items, respectively, and use the remaining interactions as the training set.
	We define interaction longer than fifty percent of the video duration as positive feedback and interaction shorter than three seconds as passive-negative feedback, following the commonly accepted industrial experience. We have also tried other definitions of positive and passive-negative feedback, and the improvement of our proposed method still holds.
	
\item \textbf{Kuaishou\footnote{\url{https://www.kuaishou.com}}.} 
This large-scale dataset is collected from Kuaishou, one of the most famous short-video platforms. 
\revise{There are billions of active users on Kuaishou, with various types of short videos, ranging from movie clips to news, as well as videos uploaded by users themselves.}
We utilize the behavioral logs collected from October 22 to October 28, 2020, with a one-week period. For the Kuaishou dataset, we have the same data pre-processing as the WeChat dataset.
\end{itemize}

\subsubsection{Metrics}
To evaluate our model and baseline models, we adopt three widely-used metrics, including AUC, GAUC, and NDCG, defined as follows.
\begin{itemize}[leftmargin=*]
	\item GAUC~\cite{zhou2018deep} is an improved version of AUC, which alleviates the negative impact of unbalanced distributions across users. It evaluates whether the model can well rank positive items higher than negative items. 
	\item NDCG is an improved version of Recall which evaluates whether the model can well rank positive items at top-K positions, and it also considers the specific position by assigning weights to the scores. In our experiments, we set $K=2$ following existing works~\cite{chang2021sequential}.
\end{itemize}
\subsubsection{Baselines}

We compare the proposed SINE with the following competitive recommenders to evaluate the performance.
\begin{itemize}[leftmargin=*]
    \item \textbf{SASRec}~\cite{kang2018self}: This model is the state-of-the-art sequential recommendation model with self-attention layers to capture sequential preferences as context vectors.

    \item \textbf{SLi-Rec}~\cite{yu2019adaptive}: This method deploys two encoders to capture long-term and short-term preferences.
    \item \textbf{DIN}~\cite{zhou2018deep}: This method proposes an attention network to obtain the similarity between historical items and the target item, to calculate the interaction probability.
    \item \textbf{DIEN}~\cite{zhou2019deep}: This method extends DIN by combining a recurrent neural network. 
    \item \textbf{CASER}~\cite{tang2018personalized}: This method adopts convolutional filters to extract the sequential patterns in user behaviors.
    \item \textbf{GRU4REC}~\cite{hidasi2015session}: This method utilizes a GRU network for modeling the users' sequential interactions.

 \item \textbf{CLSR}~\cite{zheng2022disentangling}: This method extends SLi-Rec based on disentangled representation learning, showing the state-of-the-art performance.
\item \textbf{FeedRec}~\cite{wu2022feedrec}: This model leverages various types of feedback in the sequential recommendation, and the negative feedback can be roughly treated as one kind of feedback to adapt this model to our problem.
\item \textbf{SASRec-N}~\cite{kang2018self}: Although SASRec is not defined for exploiting truly negative feedback, we can still adapt it with negative sampling from both passive-negative feedback and unobserved items. We name it SASRec-N (N denotes Negative).
\end{itemize}

It is worth mentioning FeedRec~\cite{wu2022feedrec} can be regarded as a kind of multi-feedback learning, which we will discuss in detail in Section~\ref{sec::related}. We do not include other existing methods of multi-feedback learning since they mainly focus on collaborative filtering \revise{and cannot process the sequential information.}

\subsubsection{Hyper-parameter Settings}
We implement our SINE model and the baselines on the Microsoft Recommenders framework~\cite{argyriou2020microsoft}.
 We use the Adam optimizer~\cite{kingma2014adam}, carefully searching the choice of learning rate among \{0.0005, 0.0007, 0.0009, 0.001, 0.003, 0.005\}. The batch size is tuned among \{32, 64, 128, 256, 512, 1024\}.
The embedding size $D$ of all the models is set as $50$ following existing work~\cite{kang2018self} to ensure fair performance comparison.
Besides, we find setting the number of sub-interests to two or seven for our SINE model can both achieve good recommendation performance.
We use grid-search to find the best hyper-parameters carefully, and we have released code and the best settings of hyper-parameters in  \textit{\url{https://github.com/tsinghua-fib-lab/RecSys2023-SINE}}, to benefit the community.

\subsection{Overall Performance Comparison (RQ1)}
We present the overall performance comparison on our adopted datasets in Table~\ref{tab::overall}.
Based on the results, we have the following conclusions:
\begin{itemize}[leftmargin=*]
	\item \textbf{Our SINE achieves steady and significant improvement compared with other methods.} On the Kuaishou dataset, SINE outperforms the best baseline, SASRec, by 5.32\%, 3.95\%, and 1.88\%, on AUC, GAUC, and NDCG, respectively. On the WeChat dataset, SINE outperforms the best baseline, SASRec, by 1.56\%, average on AUC, GAUC, and NDCG. 
 \revise{It is acknowledged 1\%-level improvements in AUC, GAUC, and NDCG@2 can be claimed as significant~\cite{wu2022feedrec, zheng2022disentangling, zhou2019deep}. }

	\item \textbf{The modeling of negative feedback should be carefully designed.} FeedRec and SASRec-N are two models that can utilize the negative feedback in modeling, but however, FeedRec performances worse than SASRec-N. This can be explained that the negative feedback may worsen the recommendation performance without proper utilization manner. Specifically, the observed passive-negative behaviors reflect weaker ``dislike'' signals compared with active-negative ones.
 \item \textbf{Only modeling positive feedback may achieve very poor performance.} Some competitive methods, such as CLSR and SLi-Rec, have shown good performance in datasets of traditional sequential recommendation. However, these methods perform poorly in Kuaishou and WeChat datasets since that positive feedback is far sparser in the short-video recommendation. This observation further supports this work's motivation of modeling the passive-negative feedback.
\end{itemize}

\begin{table*}[t!]
    \caption{Overall performance comparison of our SINE model and the baseline models on two datasets (the best performance is marked with bold font, and the best baseline is marked with underline).}\label{tab::overall}
    \centering
    \begin{tabular}{ccccccc}
        \toprule
        \multirow{2}{*}{\textbf{Method}} & \multicolumn{3}{c}{\textbf{Kuaishou}} & \multicolumn{3}{c}{\textbf{WeChat}} \\
        \cmidrule(l{2pt}r{2pt}){2-4}
        \cmidrule(l{2pt}r{2pt}){5-7}
    & AUC & GAUC & NDCG@2 & AUC & GAUC & NDCG@2 \\
        \midrule
        SASRec & 0.5830 & 0.5305 & \underline{0.7002} & \underline{0.6454} & \underline{0.5408} & \underline {0.6654} \\
        CLSR & 0.5676 & \underline{0.5617} & 0.6428 & 0.6423 & 0.5325 & 0.5573\\
        SLi-Rec & 0.5350 & 0.5252 & 0.6177 & 0.5901 & 0.5361 & 0.5429 \\
        DIN & \underline{0.6111} & 0.5216 & 0.4766 & 0.6379 & 0.5173 & 0.4557 \\
        GRU4REC & 0.5483 & 0.5278 & 0.5847 & 0.6193 & 0.5251 & 0.551 \\
        DIEN & 0.5840 & 0.5346 & 0.5917 & 0.6497 & 0.5361 & 0.5413 \\
        CASER & 0.5617 & 0.5260 & 0.6136 & 0.5969 & 0.5306 & 0.5496 \\
        NCF & 0.5049 & 0.5221 & 0.5961 & 0.6203 & 0.5301 & 0.5434 \\
        SASRec-N & 0.5509 & 0.5200 & 0.6958 & 0.5679 & 0.5245 & 0.6622 \\
        FeedRec & 0.5367 & 0.5102 & 0.6922 & 0.5470 & 0.5249 & 0.6612 \\
        \midrule
        Our SINE & \bf 0.6362 & \bf 0.5700 & \bf 0.7190 & \bf 0.6609 & \bf 0.5623 & \bf 0.6752 \\
        \bottomrule
    \end{tabular}
    \end{table*}

\subsection{Ablation Study (RQ2)}
In this section, we conduct experiments to study the impact of some key components, including negative feedback learning and adaptive fusion.
We show the performance results in Table~\ref{tbl:ablation}, where ``AF'' denotes the adaptive fusion-based prediction and ``NF'' denotes learning from negative feedback.

\subsubsection{Adaptive fusion}
We remove the adaptive fusion, and then there is always one specific sub-interest activated in determining user behavior. The result shows a performance drop of 0.49-5.56\%, and thus it is essential to well handle the role of different sub-interests since each behavior may be affected by multiple sub-interests.

\subsubsection{Negative feedback learning}
We remove all the negative feedback learning in our SINE model, and then the sub-interest design will not exist. It means that the SINE model will degenerate into a basic model that only leverages positive feedback. The result shows a performance drop of 1.21-6.42\%, and it illustrates the significance of our proposed whole framework of sub-interest-based negative feedback learning.

 \begin{table*}[t!]
	\caption{Ablation study of our proposed components. ``AF'' denotes adaptive-fusion-based prediction, and ``NF'' denotes learning from negative feedback.} \label{tbl:ablation}
	\centering
        \begin{tabular}{ccccccc}
            \toprule
            \multirow{2}{*}{\textbf{Method}} & \multicolumn{3}{c}{\textbf{Kuaishou}} & \multicolumn{3}{c}{\textbf{WeChat}} \\
            \cmidrule(l{2pt}r{2pt}){2-4}
            \cmidrule(l{2pt}r{2pt}){5-7}
            & AUC & GAUC & NDCG & AUC & GAUC & NDCG \\
            \midrule
            \textbf{w/o AF} & 0.6204 & 0.5411 & 0.7055 & 0.6053 & 0.5483 & 0.6703 \\
            \textbf{w/o NF} & 0.5720 & 0.5295 & 0.6988 & 0.6076 & 0.5293 & 0.6631 \\
            \textbf{Our SINE} & \textbf{0.6362} & \textbf{0.5700} & \textbf{0.7190} & \textbf{0.6609} & \textbf{0.5623} & \textbf{0.6752} \\
            \bottomrule
        \end{tabular}
\end{table*}

\begin{figure*}[t!]
	\centering
	\subfigure[GAUC w.r.t sub-interest on WeChat.]{
		\includegraphics[width=0.22\linewidth]{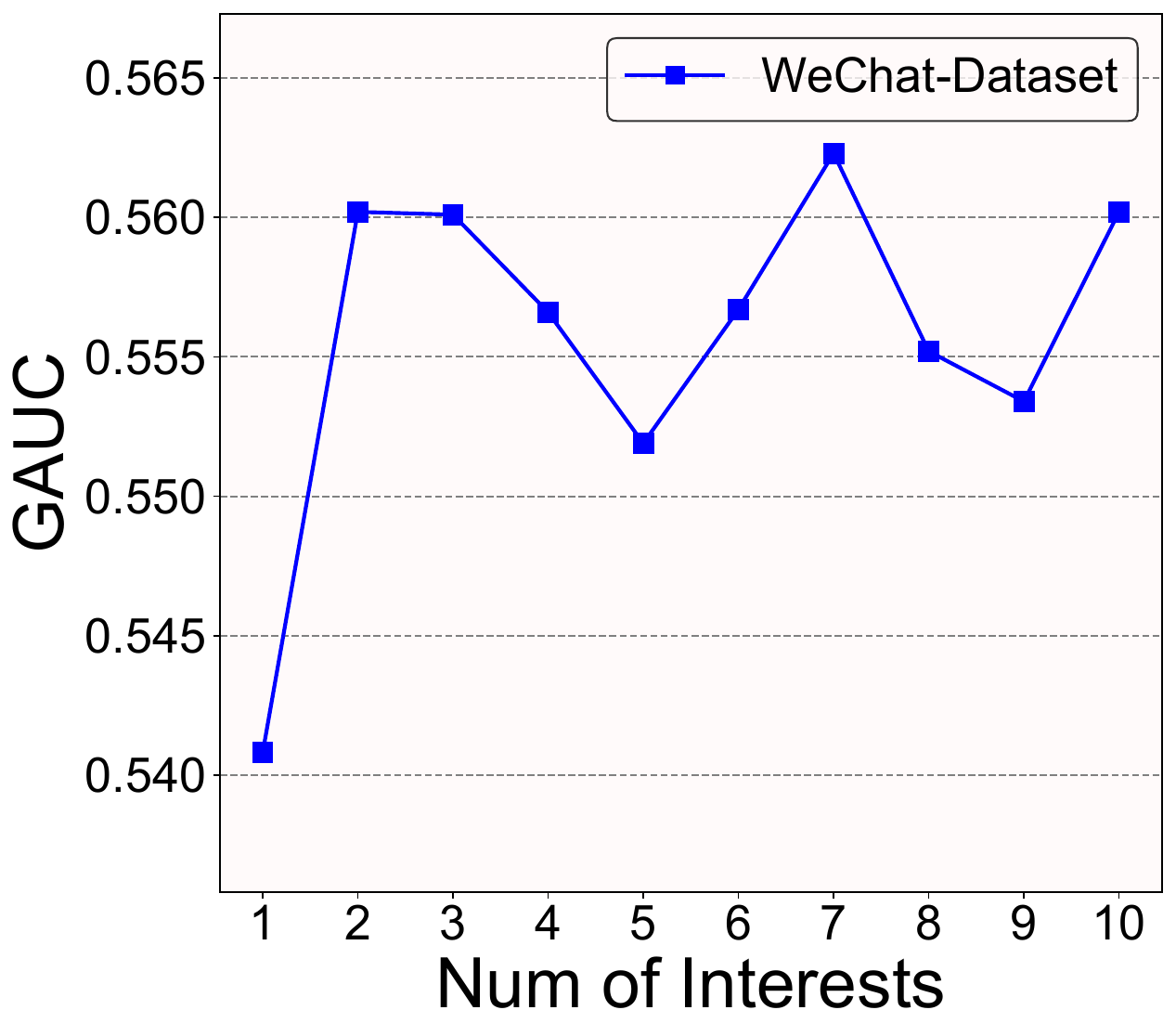}
	}
	\subfigure[NDCG w.r.t sub-interest on WeChat.]{
		\includegraphics[width=0.22\linewidth]{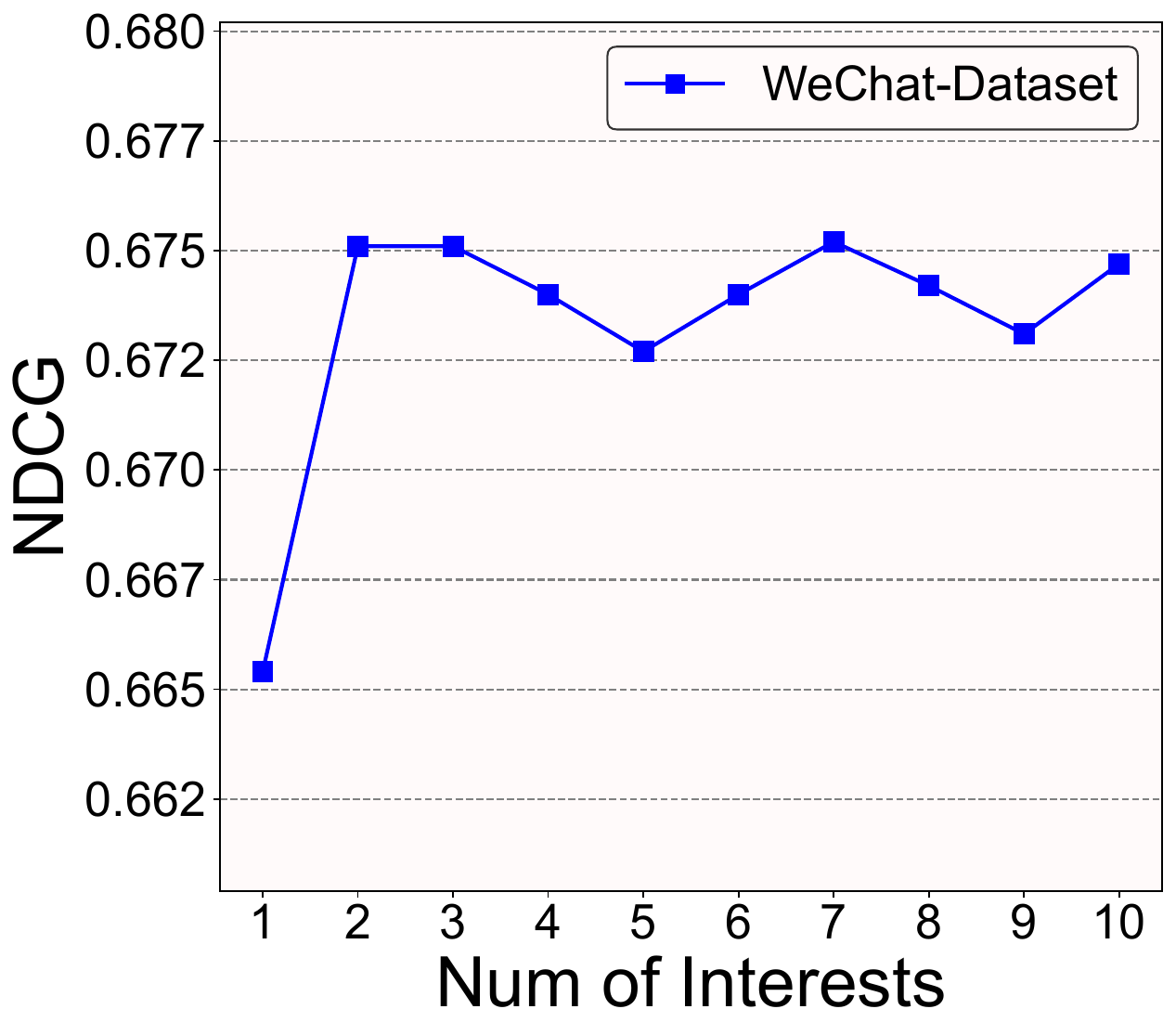}
	}
	\subfigure[GAUC w.r.t sub-interest on Kuaishou.]{
		\includegraphics[width=0.22\linewidth]{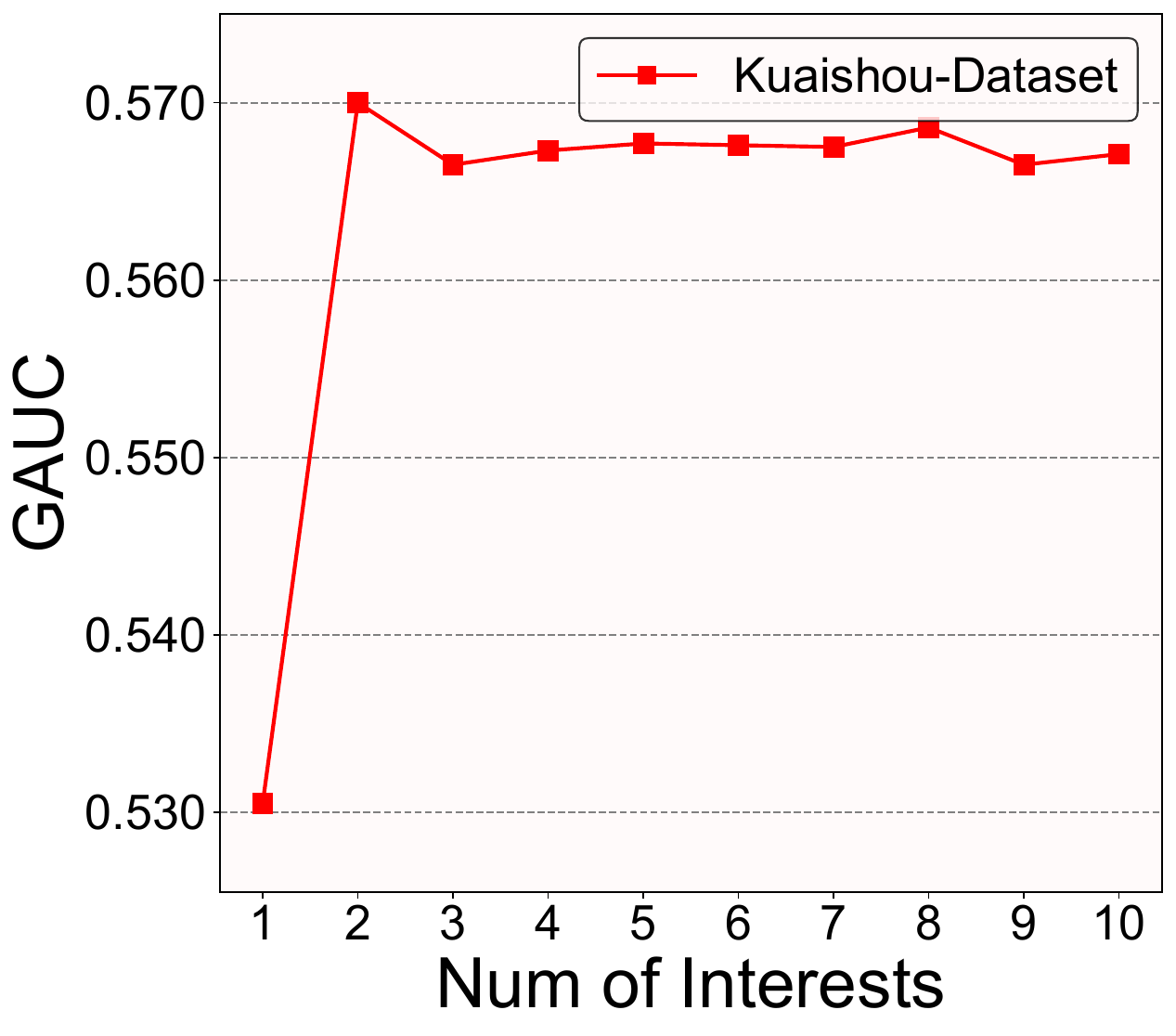}
	}
	\subfigure[NDCG w.r.t sub-interest on Kuaishou.]{
		\includegraphics[width=0.22\linewidth]{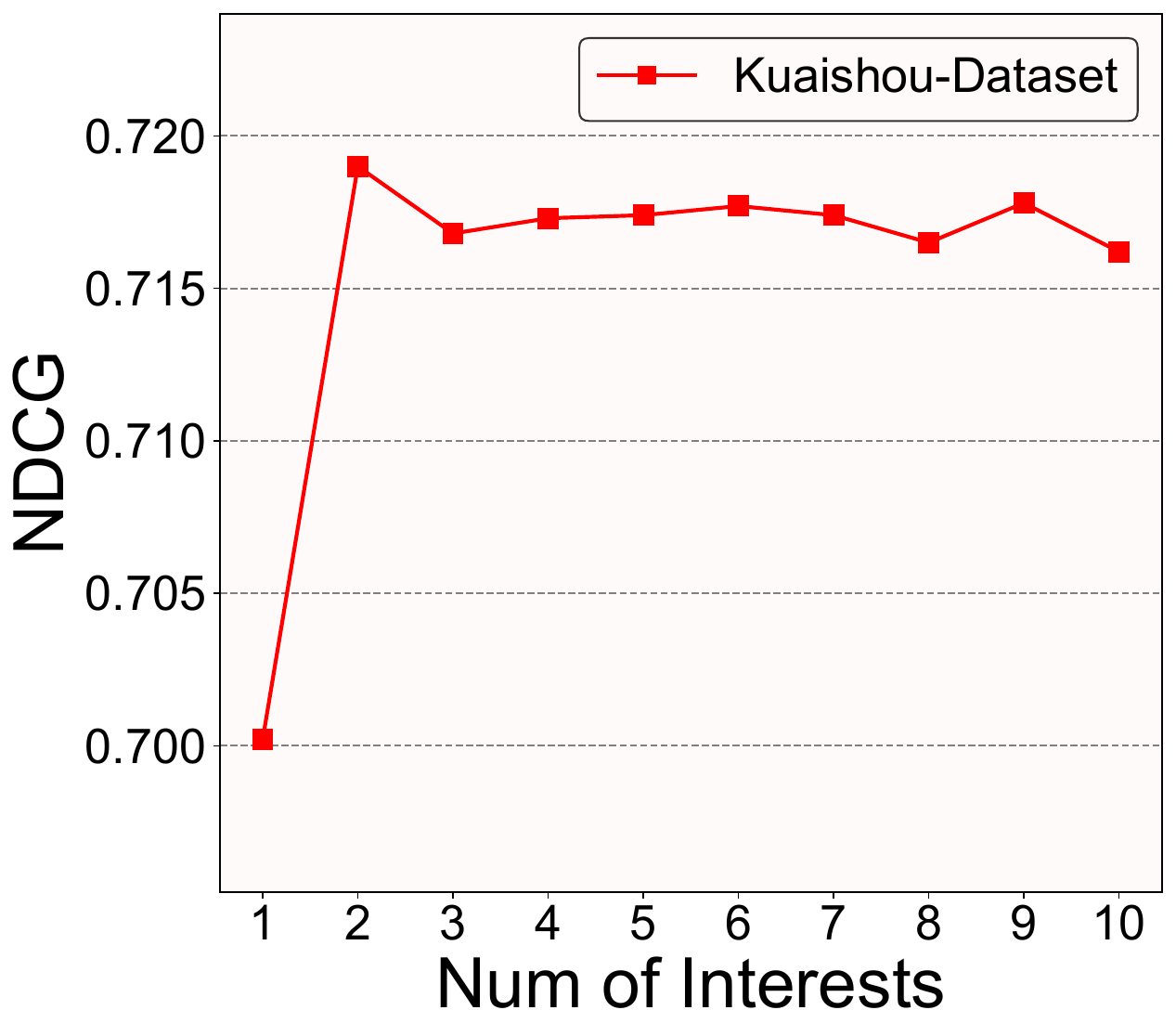}
	}
	\caption{The impact of the number of user interests on recommendation performance of our SINE model.}
	\label{fig:interest}
\end{figure*}

\subsection{Hyper-parameter Study (RQ3)}
In this section, we conduct experiments  on two datasets to study the impact of three important hyper-parameters in our SINE model, including the number of user interests, learning rate, and batch size.
\subsubsection{The number of sub-interests}
The number of sub-interests is an important hyper-parameter in our SINE model.
To explore its impact on our model's performance, we conduct experiments on the number of sub-interests in the range of \{1, 2, 3, 4, 5, 6, 7, 8, 9, 10\}. The results on two datasets are displayed in Fig.~\ref{fig:interest}, from which we have the following observations.
\begin{itemize}[leftmargin=*]
    \item \textbf{Our proposed sub-interest-based encoder is essential and effective in modeling user preference.} 
    \revise{From the results in Fig.~\ref{fig:interest}, there is a significant performance increase in models that capture two sub-interests compared to models that only capture one sub-interest. This phenomenon can be observed in both datasets across all metrics.}
    As previously discussed, we believe that users have multiple sub-interests, but only one sub-interest is active during a certain period, while the rest of the sub-interests are relatively inactive. Our proposed method utilizes users' passive-negative feedback to distinguish which sub-interest is the current active one. 
    When there is only one sub-interest, our model is regressed to the original SASRec model. However, with only two sub-interests being used to capture users' preferences, our model achieves distinct improvements compared with one sub-interest, which verifies the effectiveness of our proposed approach.
    \item \textbf{The number of user interests is customized to specific dataset.}
    The results indicate that the optimal number of sub-interests for modeling the Kuaishou dataset is two and for the WeChat dataset is seven. As shown in Fig.~\ref{fig:interest}, the overall performance of both datasets exhibits a similar trend. However, upon analyzing each dataset individually, we can observe that the best results are obtained at different numbers of sub-interests. These results demonstrate that different platforms possess unique characteristics.
\end{itemize}
\subsubsection{Learning rate \& Batch size}
We carefully tune the learning rate in the range of \{0.0005, 0.0007, 0.0009, 0.001, 0.003, 0.005\} and batch size in the range of \{32, 64, 128, 256, 512, 1024\}, following the existing works. 
Our experiments revealed that the best performance was achieved with a learning rate of 0.003 and a batch size of 32.

\vspace{0.2cm}
To summarize, we conduct experiments on two large-scale real-world datasets, and the results show our SINE's better performance compared with SOTA models.
Further experiments of ablation study verify the rationality of our model design. 

\section{Related Work}\label{sec::related}

In this section, we would like to discuss the related works based on the following three perspectives, including
sequential recommendation, multi-feedback learning in recommendation, and sub-interest learning for recommendation.
We emphasize why these methods cannot well address the studied new problem in this paper. 

\subsection{Sequential Recommendation}
Sequential recommendation is defined as recommending the next interacted item based on the historically-interacted item sequence.
The early non-deep-learning approaches~\cite{rendle2010factorizing} used the Markov chain to model the transition between items in one sequence.
Recently, deep learning methods~\cite{hidasi2015session, zhu2017next, tang2018personalized, zhou2019deep, kang2018self, sun2019bert4rec,li2017neural,li2020time,ma2020memory, gao2022causal} have become the mainstream solution for sequential recommendation.
For example, Kang~\textit{et al.~\cite{kang2018self}} build self-attention layers to encode the users' sequence.

\vspace{0.15cm}

Some recent works~\cite{an2019neural, hu2020graph, zheng2022disentangling} argued that existing sequential recommenders tend to focus on users' recent interactions and ignore the long-term user behaviors, and propose various solutions for modeling both long-term and short-term user preferences.
In industry, sequential recommenders in the large-scale system are also widely concerned~\cite{pi2020search,cao2022sampling}. 
Pi~\textit{et al.}~\cite{pi2020search} propose to use search behavior to retrieve a similar item, addressing the challenge of modeling too-long sequences. Cao~\textit{et al.}~\cite{cao2022sampling} further propose new sampling strategies which can shorten the original long user behavioral sequences.

\vspace{0.2cm}

However, these works of sequential recommendation always mainly model the sequence of positive feedback, ignoring the widely-existed negative feedback. This could be explained by the fact that most of the existing works make use of the e-commerce datasets; however, in the recent applications of short-video recommendation, the negative behavior (especially for the passive skipping-over behavior) widely exists and is very important for modeling user preferences, which is the focus of this paper.

\subsection{Multi-feedback learning in recommendation}
It is not new for recommender systems to utilize multiple types of behaviors.
Multi-behavior recommendation~\cite{jin2020multi,xia2021graph, gao2023survey} is one of the relevant research problems, which is defined as leveraging the multiple types of user feedback, such as click, adding-to-cart, purchase, etc., in e-commerce websites or click, like, share, etc. in online social media.
For example, Jin~\textit{et al.}~\cite{jin2020multi} build a multi-relational user-item graph to represent the multi-form feedback between users and items, and then developed graph neural network models to predict the missing links on the graph.
That is, these works tend to study the multi-type positive feedback, leaving negative feedback less explored.

Another close topic is exposure bias-aware recommendation~\cite{gupta2021correcting,mansoury2021graph,chen2021autodebias}, in which the expose-but-not-click is also considered as a kind of feedback, and it is related to the passive-negative feedback in this work.
Other works approach the implicit feedback modeling by designing a negative sampling strategy via exposure data~\cite{ding2019reinforced, ding2020simplify}.
The recent negative feedback modeling works \cite{wang2022unbiased, gong2022positive, seo2022siren, DBLP:conf/ijcai/XieLWWXL20} also explore this expose-but-not-click negative feedback.
Wang~\textit{et al.}\cite{wang2022unbiased} focuses on unbiased recommendation that introduces negative feedback modeling and Seo~\textit{et al.}\cite{seo2022siren} is a graph-based recommendation.

However, there are two folds of critical differences compared with our work. First, the user may have not noticed the exposed items as the utilized datasets in these works still require the user to actively click. Second, the sequential behaviors are not well considered in these works.

\subsection{Sub-interest learning for recommendation}
Users in the real world always have multiple criteria to judge whether the recommended item satisfies their needs or not.
For example, a user of an e-commerce website will make decisions according to price, brand, functionality, etc.
Thus, different from the traditional recommendation models with only one general interest representation~\cite{he2017neural},
researchers have begun to model multiple distinct interests of users, which are also known as sub-interests~\cite{li2019multi,chai2022user,sabour2017dynamic,chen2021multi,tian2022multi}. 

Li~\textit{et al.}~\cite{li2019multi} proposed to use capsule network~\cite{sabour2017dynamic} to extract the users' sub-interests and combined it with a label-aware attention layer.
Chai~\textit{et al.}~\cite{chai2022user} combines capsule network with user profile information to refine the user’s sub-interest in enhancing recommendations.
Cen~\textit{et al.}~\cite{cen2020controllable} propose two kinds of architectures, capsule network, and self-attention network, with a controllable procedure to balance the accuracy and diversity of recommendation.
Similarly, Chen~\textit{et al.}~\cite{chen2021multi}also extracts sub-interests with capsule network and self-attention network.
Recently, Tian~\textit{et al.}~\cite{tian2022multi} propose to combine multi-interest learning with multi-grained interest learning.

However, the sub-interests of these works lack explicit supervision signals, and the learned sub-interests are always not explainable.
Different from them, in our work, we use the sub-interests to explain the occurrence of passive-negative feedback, which can serve as good supervision for the representation learning of sub-interests.

\section{Conclusion and Future Work}\label{sec::conclusion}
In this work, we approach a new problem in sequential recommender systems: understanding and modeling users' passive-negative feedback in short-video sequential recommendation.
We first use the data analysis on large-scale real-world data to demonstrate 1) it is important,  2) it is challenging, and 3) why it is challenging to utilize passive-negative feedback in sequential recommenders.
\revise{The results show that the passive-negative feedback is similar to the observed positive feedback in the video category and thus the intuitive manner of treating it as a negative sample does not work well.}
Motivated by the data analysis, we propose a method with a sub-interest extractor, in which the passive-negative feedback can be modeled as the mismatch of specific sub-interests.
The performance comparison shows the best performance of our proposed SINE method.
The further results extensively demonstrate the effectiveness of SINE's different components, and the results correspond well to the earlier data analysis.

As for future work, we first plan to deploy the proposed SINE model to the real-world online recommendation engine and evaluate the recommendation performance compared with other methods through the A/B tests. We also plan to collect other kinds of explicit positive feedback, such as like behavior, sharing behavior, etc., to enhance the study of negative feedback in recommender systems, which can motivate us to further improve the model design.

\begin{acks}
This work is partially supported by the National Key Research and Development Program of China under 2022YFB3104702, and the National Natural Science Foundation of China under 62272262, 61972223, U1936217, and U20B2060.
\end{acks}

\balance
\bibliographystyle{ACM-Reference-Format}
\bibliography{sample-base}


\begin{thebibliography}{49}


\ifx \showCODEN    \undefined \def \showCODEN     #1{\unskip}     \fi
\ifx \showDOI      \undefined \def \showDOI       #1{#1}\fi
\ifx \showISBNx    \undefined \def \showISBNx     #1{\unskip}     \fi
\ifx \showISBNxiii \undefined \def \showISBNxiii  #1{\unskip}     \fi
\ifx \showISSN     \undefined \def \showISSN      #1{\unskip}     \fi
\ifx \showLCCN     \undefined \def \showLCCN      #1{\unskip}     \fi
\ifx \shownote     \undefined \def \shownote      #1{#1}          \fi
\ifx \showarticletitle \undefined \def \showarticletitle #1{#1}   \fi
\ifx \showURL      \undefined \def \showURL       {\relax}        \fi
\providecommand\bibfield[2]{#2}
\providecommand\bibinfo[2]{#2}
\providecommand\natexlab[1]{#1}
\providecommand\showeprint[2][]{arXiv:#2}

\bibitem[\protect\citeauthoryear{An, Wu, Wu, Zhang, Liu, and Xie}{An
  et~al\mbox{.}}{2019}]%
        {an2019neural}
\bibfield{author}{\bibinfo{person}{Mingxiao An}, \bibinfo{person}{Fangzhao Wu},
  \bibinfo{person}{Chuhan Wu}, \bibinfo{person}{Kun Zhang},
  \bibinfo{person}{Zheng Liu}, {and} \bibinfo{person}{Xing Xie}.}
  \bibinfo{year}{2019}\natexlab{}.
\newblock \showarticletitle{Neural news recommendation with long-and short-term
  user representations}. In \bibinfo{booktitle}{\emph{Proceedings of the 57th
  Annual Meeting of the Association for Computational Linguistics}}.
  \bibinfo{pages}{336--345}.
\newblock


\bibitem[\protect\citeauthoryear{Argyriou, Gonz{\'a}lez-Fierro, and
  Zhang}{Argyriou et~al\mbox{.}}{2020}]%
        {argyriou2020microsoft}
\bibfield{author}{\bibinfo{person}{Andreas Argyriou}, \bibinfo{person}{Miguel
  Gonz{\'a}lez-Fierro}, {and} \bibinfo{person}{Le Zhang}.}
  \bibinfo{year}{2020}\natexlab{}.
\newblock \showarticletitle{Microsoft Recommenders: Best Practices for
  Production-Ready Recommendation Systems}. In
  \bibinfo{booktitle}{\emph{Companion Proceedings of the Web Conference 2020}}.
  \bibinfo{pages}{50--51}.
\newblock


\bibitem[\protect\citeauthoryear{Cao, Zhou, Feng, Huang, Xiao, Chen, and
  Chen}{Cao et~al\mbox{.}}{2022}]%
        {cao2022sampling}
\bibfield{author}{\bibinfo{person}{Yue Cao}, \bibinfo{person}{XiaoJiang Zhou},
  \bibinfo{person}{Jiaqi Feng}, \bibinfo{person}{Peihao Huang},
  \bibinfo{person}{Yao Xiao}, \bibinfo{person}{Dayao Chen}, {and}
  \bibinfo{person}{Sheng Chen}.} \bibinfo{year}{2022}\natexlab{}.
\newblock \showarticletitle{Sampling Is All You Need on Modeling Long-Term User
  Behaviors for CTR Prediction}. In \bibinfo{booktitle}{\emph{CIKM}}.
\newblock


\bibitem[\protect\citeauthoryear{Cen, Zhang, Zou, Zhou, Yang, and Tang}{Cen
  et~al\mbox{.}}{2020}]%
        {cen2020controllable}
\bibfield{author}{\bibinfo{person}{Yukuo Cen}, \bibinfo{person}{Jianwei Zhang},
  \bibinfo{person}{Xu Zou}, \bibinfo{person}{Chang Zhou},
  \bibinfo{person}{Hongxia Yang}, {and} \bibinfo{person}{Jie Tang}.}
  \bibinfo{year}{2020}\natexlab{}.
\newblock \showarticletitle{Controllable multi-interest framework for
  recommendation}. In \bibinfo{booktitle}{\emph{Proceedings of the 26th ACM
  SIGKDD International Conference on Knowledge Discovery \& Data Mining}}.
  \bibinfo{pages}{2942--2951}.
\newblock


\bibitem[\protect\citeauthoryear{Chai, Chen, Li, Xiao, Li, Wu, Chen, and
  Tang}{Chai et~al\mbox{.}}{2022}]%
        {chai2022user}
\bibfield{author}{\bibinfo{person}{Zheng Chai}, \bibinfo{person}{Zhihong Chen},
  \bibinfo{person}{Chenliang Li}, \bibinfo{person}{Rong Xiao},
  \bibinfo{person}{Houyi Li}, \bibinfo{person}{Jiawei Wu},
  \bibinfo{person}{Jingxu Chen}, {and} \bibinfo{person}{Haihong Tang}.}
  \bibinfo{year}{2022}\natexlab{}.
\newblock \showarticletitle{User-Aware Multi-Interest Learning for Candidate
  Matching in Recommenders}. In \bibinfo{booktitle}{\emph{Proceedings of the
  45th International ACM SIGIR Conference on Research and Development in
  Information Retrieval}}. \bibinfo{pages}{1326--1335}.
\newblock


\bibitem[\protect\citeauthoryear{Chang, Gao, Zheng, Hui, Niu, Song, Jin, and
  Li}{Chang et~al\mbox{.}}{2021}]%
        {chang2021sequential}
\bibfield{author}{\bibinfo{person}{Jianxin Chang}, \bibinfo{person}{Chen Gao},
  \bibinfo{person}{Yu Zheng}, \bibinfo{person}{Yiqun Hui},
  \bibinfo{person}{Yanan Niu}, \bibinfo{person}{Yang Song},
  \bibinfo{person}{Depeng Jin}, {and} \bibinfo{person}{Yong Li}.}
  \bibinfo{year}{2021}\natexlab{}.
\newblock \showarticletitle{Sequential Recommendation with Graph Neural
  Networks}. In \bibinfo{booktitle}{\emph{Proceedings of the 44th International
  ACM SIGIR Conference on Research and Development in Information Retrieval}}.
  \bibinfo{pages}{378--387}.
\newblock


\bibitem[\protect\citeauthoryear{Chen, Dong, Qiu, He, Xin, Chen, Lin, and
  Yang}{Chen et~al\mbox{.}}{2021a}]%
        {chen2021autodebias}
\bibfield{author}{\bibinfo{person}{Jiawei Chen}, \bibinfo{person}{Hande Dong},
  \bibinfo{person}{Yang Qiu}, \bibinfo{person}{Xiangnan He},
  \bibinfo{person}{Xin Xin}, \bibinfo{person}{Liang Chen},
  \bibinfo{person}{Guli Lin}, {and} \bibinfo{person}{Keping Yang}.}
  \bibinfo{year}{2021}\natexlab{a}.
\newblock \showarticletitle{AutoDebias: Learning to debias for recommendation}.
  In \bibinfo{booktitle}{\emph{Proceedings of the 44th International ACM SIGIR
  Conference on Research and Development in Information Retrieval}}.
  \bibinfo{pages}{21--30}.
\newblock


\bibitem[\protect\citeauthoryear{Chen, Ren, Cai, Sun, and De~Rijke}{Chen
  et~al\mbox{.}}{2021b}]%
        {chen2021multi}
\bibfield{author}{\bibinfo{person}{Wanyu Chen}, \bibinfo{person}{Pengjie Ren},
  \bibinfo{person}{Fei Cai}, \bibinfo{person}{Fei Sun}, {and}
  \bibinfo{person}{Maarten De~Rijke}.} \bibinfo{year}{2021}\natexlab{b}.
\newblock \showarticletitle{Multi-interest diversification for end-to-end
  sequential recommendation}.
\newblock \bibinfo{journal}{\emph{ACM Transactions on Information Systems
  (TOIS)}} \bibinfo{volume}{40}, \bibinfo{number}{1} (\bibinfo{year}{2021}),
  \bibinfo{pages}{1--30}.
\newblock


\bibitem[\protect\citeauthoryear{Cho, Hyun, Kang, and Yu}{Cho
  et~al\mbox{.}}{2021}]%
        {cho2021learning}
\bibfield{author}{\bibinfo{person}{Junsu Cho}, \bibinfo{person}{Dongmin Hyun},
  \bibinfo{person}{Seongku Kang}, {and} \bibinfo{person}{Hwanjo Yu}.}
  \bibinfo{year}{2021}\natexlab{}.
\newblock \showarticletitle{Learning heterogeneous temporal patterns of user
  preference for timely recommendation}. In
  \bibinfo{booktitle}{\emph{Proceedings of the Web Conference 2021}}.
  \bibinfo{pages}{1274--1283}.
\newblock


\bibitem[\protect\citeauthoryear{Ding, Quan, He, Li, and Jin}{Ding
  et~al\mbox{.}}{2019}]%
        {ding2019reinforced}
\bibfield{author}{\bibinfo{person}{Jingtao Ding}, \bibinfo{person}{Yuhan Quan},
  \bibinfo{person}{Xiangnan He}, \bibinfo{person}{Yong Li}, {and}
  \bibinfo{person}{Depeng Jin}.} \bibinfo{year}{2019}\natexlab{}.
\newblock \showarticletitle{Reinforced Negative Sampling for Recommendation
  with Exposure Data.}. In \bibinfo{booktitle}{\emph{IJCAI}}. Macao,
  \bibinfo{pages}{2230--2236}.
\newblock


\bibitem[\protect\citeauthoryear{Ding, Quan, Yao, Li, and Jin}{Ding
  et~al\mbox{.}}{2020}]%
        {ding2020simplify}
\bibfield{author}{\bibinfo{person}{Jingtao Ding}, \bibinfo{person}{Yuhan Quan},
  \bibinfo{person}{Quanming Yao}, \bibinfo{person}{Yong Li}, {and}
  \bibinfo{person}{Depeng Jin}.} \bibinfo{year}{2020}\natexlab{}.
\newblock \showarticletitle{Simplify and robustify negative sampling for
  implicit collaborative filtering}.
\newblock \bibinfo{journal}{\emph{Advances in Neural Information Processing
  Systems}}  \bibinfo{volume}{33} (\bibinfo{year}{2020}),
  \bibinfo{pages}{1094--1105}.
\newblock


\bibitem[\protect\citeauthoryear{Gao, Zheng, Li, Li, Qin, Piao, Quan, Chang,
  Jin, He, et~al\mbox{.}}{Gao et~al\mbox{.}}{2023}]%
        {gao2023survey}
\bibfield{author}{\bibinfo{person}{Chen Gao}, \bibinfo{person}{Yu Zheng},
  \bibinfo{person}{Nian Li}, \bibinfo{person}{Yinfeng Li},
  \bibinfo{person}{Yingrong Qin}, \bibinfo{person}{Jinghua Piao},
  \bibinfo{person}{Yuhan Quan}, \bibinfo{person}{Jianxin Chang},
  \bibinfo{person}{Depeng Jin}, \bibinfo{person}{Xiangnan He}, {et~al\mbox{.}}}
  \bibinfo{year}{2023}\natexlab{}.
\newblock \showarticletitle{A survey of graph neural networks for recommender
  systems: challenges, methods, and directions}.
\newblock \bibinfo{journal}{\emph{ACM Transactions on Recommender Systems}}
  \bibinfo{volume}{1}, \bibinfo{number}{1} (\bibinfo{year}{2023}),
  \bibinfo{pages}{1--51}.
\newblock


\bibitem[\protect\citeauthoryear{Gao, Zheng, Wang, Feng, He, and Li}{Gao
  et~al\mbox{.}}{2022}]%
        {gao2022causal}
\bibfield{author}{\bibinfo{person}{Chen Gao}, \bibinfo{person}{Yu Zheng},
  \bibinfo{person}{Wenjie Wang}, \bibinfo{person}{Fuli Feng},
  \bibinfo{person}{Xiangnan He}, {and} \bibinfo{person}{Yong Li}.}
  \bibinfo{year}{2022}\natexlab{}.
\newblock \showarticletitle{Causal Inference in Recommender Systems: A Survey
  and Future Directions}.
\newblock \bibinfo{journal}{\emph{arXiv preprint arXiv:2208.12397}}
  (\bibinfo{year}{2022}).
\newblock


\bibitem[\protect\citeauthoryear{Gong and Zhu}{Gong and Zhu}{2022}]%
        {gong2022positive}
\bibfield{author}{\bibinfo{person}{Shansan Gong} {and} \bibinfo{person}{Kenny~Q
  Zhu}.} \bibinfo{year}{2022}\natexlab{}.
\newblock \showarticletitle{Positive, Negative and Neutral: Modeling Implicit
  Feedback in Session-based News Recommendation}. In
  \bibinfo{booktitle}{\emph{Proceedings of the 45th International ACM SIGIR
  Conference on Research and Development in Information Retrieval}}.
  \bibinfo{pages}{1185--1195}.
\newblock


\bibitem[\protect\citeauthoryear{Gupta, Wang, Lipton, and Wang}{Gupta
  et~al\mbox{.}}{2021}]%
        {gupta2021correcting}
\bibfield{author}{\bibinfo{person}{Shantanu Gupta}, \bibinfo{person}{Hao Wang},
  \bibinfo{person}{Zachary Lipton}, {and} \bibinfo{person}{Yuyang Wang}.}
  \bibinfo{year}{2021}\natexlab{}.
\newblock \showarticletitle{Correcting exposure bias for link recommendation}.
  In \bibinfo{booktitle}{\emph{International Conference on Machine Learning}}.
  PMLR, \bibinfo{pages}{3953--3963}.
\newblock


\bibitem[\protect\citeauthoryear{He, Liao, Zhang, Nie, Hu, and Chua}{He
  et~al\mbox{.}}{2017}]%
        {he2017neural}
\bibfield{author}{\bibinfo{person}{Xiangnan He}, \bibinfo{person}{Lizi Liao},
  \bibinfo{person}{Hanwang Zhang}, \bibinfo{person}{Liqiang Nie},
  \bibinfo{person}{Xia Hu}, {and} \bibinfo{person}{Tat-Seng Chua}.}
  \bibinfo{year}{2017}\natexlab{}.
\newblock \showarticletitle{Neural collaborative filtering}. In
  \bibinfo{booktitle}{\emph{Proceedings of the 26th international conference on
  world wide web}}. \bibinfo{pages}{173--182}.
\newblock


\bibitem[\protect\citeauthoryear{He, Zhang, Kan, and Chua}{He
  et~al\mbox{.}}{2016}]%
        {he2016fast}
\bibfield{author}{\bibinfo{person}{Xiangnan He}, \bibinfo{person}{Hanwang
  Zhang}, \bibinfo{person}{Min-Yen Kan}, {and} \bibinfo{person}{Tat-Seng
  Chua}.} \bibinfo{year}{2016}\natexlab{}.
\newblock \showarticletitle{Fast matrix factorization for online recommendation
  with implicit feedback}. In \bibinfo{booktitle}{\emph{Proceedings of the 39th
  International ACM SIGIR conference on Research and Development in Information
  Retrieval}}. \bibinfo{pages}{549--558}.
\newblock


\bibitem[\protect\citeauthoryear{Hidasi, Karatzoglou, Baltrunas, and
  Tikk}{Hidasi et~al\mbox{.}}{2015}]%
        {hidasi2015session}
\bibfield{author}{\bibinfo{person}{Bal{\'a}zs Hidasi},
  \bibinfo{person}{Alexandros Karatzoglou}, \bibinfo{person}{Linas Baltrunas},
  {and} \bibinfo{person}{Domonkos Tikk}.} \bibinfo{year}{2015}\natexlab{}.
\newblock \showarticletitle{Session-based recommendations with recurrent neural
  networks}.
\newblock \bibinfo{journal}{\emph{arXiv preprint arXiv:1511.06939}}
  (\bibinfo{year}{2015}).
\newblock


\bibitem[\protect\citeauthoryear{Hu, Li, Shi, Yang, and Shao}{Hu
  et~al\mbox{.}}{2020}]%
        {hu2020graph}
\bibfield{author}{\bibinfo{person}{Linmei Hu}, \bibinfo{person}{Chen Li},
  \bibinfo{person}{Chuan Shi}, \bibinfo{person}{Cheng Yang}, {and}
  \bibinfo{person}{Chao Shao}.} \bibinfo{year}{2020}\natexlab{}.
\newblock \showarticletitle{Graph neural news recommendation with long-term and
  short-term interest modeling}.
\newblock \bibinfo{journal}{\emph{Information Processing \& Management}}
  \bibinfo{volume}{57}, \bibinfo{number}{2} (\bibinfo{year}{2020}),
  \bibinfo{pages}{102142}.
\newblock


\bibitem[\protect\citeauthoryear{Jin, Gao, He, Jin, and Li}{Jin
  et~al\mbox{.}}{2020}]%
        {jin2020multi}
\bibfield{author}{\bibinfo{person}{Bowen Jin}, \bibinfo{person}{Chen Gao},
  \bibinfo{person}{Xiangnan He}, \bibinfo{person}{Depeng Jin}, {and}
  \bibinfo{person}{Yong Li}.} \bibinfo{year}{2020}\natexlab{}.
\newblock \showarticletitle{Multi-behavior recommendation with graph
  convolutional networks}. In \bibinfo{booktitle}{\emph{Proceedings of the 43rd
  International ACM SIGIR Conference on Research and Development in Information
  Retrieval}}. \bibinfo{pages}{659--668}.
\newblock


\bibitem[\protect\citeauthoryear{Kang and McAuley}{Kang and McAuley}{2018}]%
        {kang2018self}
\bibfield{author}{\bibinfo{person}{Wang-Cheng Kang} {and}
  \bibinfo{person}{Julian McAuley}.} \bibinfo{year}{2018}\natexlab{}.
\newblock \showarticletitle{Self-attentive sequential recommendation}. In
  \bibinfo{booktitle}{\emph{2018 IEEE International Conference on Data Mining
  (ICDM)}}. IEEE, \bibinfo{pages}{197--206}.
\newblock


\bibitem[\protect\citeauthoryear{Kingma and Ba}{Kingma and Ba}{2014}]%
        {kingma2014adam}
\bibfield{author}{\bibinfo{person}{Diederik~P Kingma} {and}
  \bibinfo{person}{Jimmy Ba}.} \bibinfo{year}{2014}\natexlab{}.
\newblock \showarticletitle{Adam: A method for stochastic optimization}.
\newblock \bibinfo{journal}{\emph{arXiv preprint arXiv:1412.6980}}
  (\bibinfo{year}{2014}).
\newblock


\bibitem[\protect\citeauthoryear{Li, Liu, Wu, Xu, Zhao, Huang, Kang, Chen, Li,
  and Lee}{Li et~al\mbox{.}}{2019}]%
        {li2019multi}
\bibfield{author}{\bibinfo{person}{Chao Li}, \bibinfo{person}{Zhiyuan Liu},
  \bibinfo{person}{Mengmeng Wu}, \bibinfo{person}{Yuchi Xu},
  \bibinfo{person}{Huan Zhao}, \bibinfo{person}{Pipei Huang},
  \bibinfo{person}{Guoliang Kang}, \bibinfo{person}{Qiwei Chen},
  \bibinfo{person}{Wei Li}, {and} \bibinfo{person}{Dik~Lun Lee}.}
  \bibinfo{year}{2019}\natexlab{}.
\newblock \showarticletitle{Multi-interest network with dynamic routing for
  recommendation at Tmall}. In \bibinfo{booktitle}{\emph{Proceedings of the
  28th ACM International Conference on Information and Knowledge Management}}.
  \bibinfo{pages}{2615--2623}.
\newblock


\bibitem[\protect\citeauthoryear{Li, Ren, Chen, Ren, Lian, and Ma}{Li
  et~al\mbox{.}}{2017}]%
        {li2017neural}
\bibfield{author}{\bibinfo{person}{Jing Li}, \bibinfo{person}{Pengjie Ren},
  \bibinfo{person}{Zhumin Chen}, \bibinfo{person}{Zhaochun Ren},
  \bibinfo{person}{Tao Lian}, {and} \bibinfo{person}{Jun Ma}.}
  \bibinfo{year}{2017}\natexlab{}.
\newblock \showarticletitle{Neural attentive session-based recommendation}. In
  \bibinfo{booktitle}{\emph{Proceedings of the 2017 ACM on Conference on
  Information and Knowledge Management}}. \bibinfo{pages}{1419--1428}.
\newblock


\bibitem[\protect\citeauthoryear{Li, Wang, and McAuley}{Li
  et~al\mbox{.}}{2020}]%
        {li2020time}
\bibfield{author}{\bibinfo{person}{Jiacheng Li}, \bibinfo{person}{Yujie Wang},
  {and} \bibinfo{person}{Julian McAuley}.} \bibinfo{year}{2020}\natexlab{}.
\newblock \showarticletitle{Time interval aware self-attention for sequential
  recommendation}. In \bibinfo{booktitle}{\emph{Proceedings of the 13th
  international conference on web search and data mining}}.
  \bibinfo{pages}{322--330}.
\newblock


\bibitem[\protect\citeauthoryear{Ma, Ma, Zhang, Sun, Liu, and Coates}{Ma
  et~al\mbox{.}}{2020}]%
        {ma2020memory}
\bibfield{author}{\bibinfo{person}{Chen Ma}, \bibinfo{person}{Liheng Ma},
  \bibinfo{person}{Yingxue Zhang}, \bibinfo{person}{Jianing Sun},
  \bibinfo{person}{Xue Liu}, {and} \bibinfo{person}{Mark Coates}.}
  \bibinfo{year}{2020}\natexlab{}.
\newblock \showarticletitle{Memory augmented graph neural networks for
  sequential recommendation}. In \bibinfo{booktitle}{\emph{Proceedings of the
  AAAI Conference on Artificial Intelligence}}, Vol.~\bibinfo{volume}{34}.
  \bibinfo{pages}{5045--5052}.
\newblock


\bibitem[\protect\citeauthoryear{Mansoury, Abdollahpouri, Pechenizkiy,
  Mobasher, and Burke}{Mansoury et~al\mbox{.}}{2021}]%
        {mansoury2021graph}
\bibfield{author}{\bibinfo{person}{Masoud Mansoury}, \bibinfo{person}{Himan
  Abdollahpouri}, \bibinfo{person}{Mykola Pechenizkiy},
  \bibinfo{person}{Bamshad Mobasher}, {and} \bibinfo{person}{Robin Burke}.}
  \bibinfo{year}{2021}\natexlab{}.
\newblock \showarticletitle{A graph-based approach for mitigating multi-sided
  exposure bias in recommender systems}.
\newblock \bibinfo{journal}{\emph{ACM Transactions on Information Systems
  (TOIS)}} \bibinfo{volume}{40}, \bibinfo{number}{2} (\bibinfo{year}{2021}),
  \bibinfo{pages}{1--31}.
\newblock


\bibitem[\protect\citeauthoryear{Pi, Zhou, Zhang, Wang, Ren, Fan, Zhu, and
  Gai}{Pi et~al\mbox{.}}{2020}]%
        {pi2020search}
\bibfield{author}{\bibinfo{person}{Qi Pi}, \bibinfo{person}{Guorui Zhou},
  \bibinfo{person}{Yujing Zhang}, \bibinfo{person}{Zhe Wang},
  \bibinfo{person}{Lejian Ren}, \bibinfo{person}{Ying Fan},
  \bibinfo{person}{Xiaoqiang Zhu}, {and} \bibinfo{person}{Kun Gai}.}
  \bibinfo{year}{2020}\natexlab{}.
\newblock \showarticletitle{Search-based user interest modeling with lifelong
  sequential behavior data for click-through rate prediction}. In
  \bibinfo{booktitle}{\emph{Proceedings of the 29th ACM International
  Conference on Information \& Knowledge Management}}.
  \bibinfo{pages}{2685--2692}.
\newblock


\bibitem[\protect\citeauthoryear{Rendle, Freudenthaler, Gantner, and
  Schmidt-Thieme}{Rendle et~al\mbox{.}}{2012}]%
        {rendle2012bpr}
\bibfield{author}{\bibinfo{person}{Steffen Rendle}, \bibinfo{person}{Christoph
  Freudenthaler}, \bibinfo{person}{Zeno Gantner}, {and} \bibinfo{person}{Lars
  Schmidt-Thieme}.} \bibinfo{year}{2012}\natexlab{}.
\newblock \showarticletitle{BPR: Bayesian personalized ranking from implicit
  feedback}.
\newblock \bibinfo{journal}{\emph{arXiv preprint arXiv:1205.2618}}
  (\bibinfo{year}{2012}).
\newblock


\bibitem[\protect\citeauthoryear{Rendle, Freudenthaler, and
  Schmidt-Thieme}{Rendle et~al\mbox{.}}{2010}]%
        {rendle2010factorizing}
\bibfield{author}{\bibinfo{person}{Steffen Rendle}, \bibinfo{person}{Christoph
  Freudenthaler}, {and} \bibinfo{person}{Lars Schmidt-Thieme}.}
  \bibinfo{year}{2010}\natexlab{}.
\newblock \showarticletitle{Factorizing personalized markov chains for
  next-basket recommendation}. In \bibinfo{booktitle}{\emph{Proceedings of the
  19th international conference on World wide web}}. \bibinfo{pages}{811--820}.
\newblock


\bibitem[\protect\citeauthoryear{Sabour, Frosst, and Hinton}{Sabour
  et~al\mbox{.}}{2017}]%
        {sabour2017dynamic}
\bibfield{author}{\bibinfo{person}{Sara Sabour}, \bibinfo{person}{Nicholas
  Frosst}, {and} \bibinfo{person}{Geoffrey~E Hinton}.}
  \bibinfo{year}{2017}\natexlab{}.
\newblock \showarticletitle{Dynamic routing between capsules}.
\newblock \bibinfo{journal}{\emph{Advances in neural information processing
  systems}}  \bibinfo{volume}{30} (\bibinfo{year}{2017}).
\newblock


\bibitem[\protect\citeauthoryear{Seo, Jeong, Lim, and Shin}{Seo
  et~al\mbox{.}}{2022}]%
        {seo2022siren}
\bibfield{author}{\bibinfo{person}{Changwon Seo}, \bibinfo{person}{Kyeong-Joong
  Jeong}, \bibinfo{person}{Sungsu Lim}, {and} \bibinfo{person}{Won-Yong Shin}.}
  \bibinfo{year}{2022}\natexlab{}.
\newblock \showarticletitle{SiReN: Sign-Aware Recommendation Using Graph Neural
  Networks}.
\newblock \bibinfo{journal}{\emph{IEEE Transactions on Neural Networks and
  Learning Systems}} (\bibinfo{year}{2022}).
\newblock


\bibitem[\protect\citeauthoryear{Sun, Liu, Wu, Pei, Lin, Ou, and Jiang}{Sun
  et~al\mbox{.}}{2019}]%
        {sun2019bert4rec}
\bibfield{author}{\bibinfo{person}{Fei Sun}, \bibinfo{person}{Jun Liu},
  \bibinfo{person}{Jian Wu}, \bibinfo{person}{Changhua Pei},
  \bibinfo{person}{Xiao Lin}, \bibinfo{person}{Wenwu Ou}, {and}
  \bibinfo{person}{Peng Jiang}.} \bibinfo{year}{2019}\natexlab{}.
\newblock \showarticletitle{BERT4Rec: Sequential recommendation with
  bidirectional encoder representations from transformer}. In
  \bibinfo{booktitle}{\emph{Proceedings of the 28th ACM International
  Conference on Information and Knowledge Management}}.
  \bibinfo{pages}{1441--1450}.
\newblock


\bibitem[\protect\citeauthoryear{Sz{\'e}kely, Rizzo, and Bakirov}{Sz{\'e}kely
  et~al\mbox{.}}{2007}]%
        {szekely2007measuring}
\bibfield{author}{\bibinfo{person}{G{\'a}bor~J Sz{\'e}kely},
  \bibinfo{person}{Maria~L Rizzo}, {and} \bibinfo{person}{Nail~K Bakirov}.}
  \bibinfo{year}{2007}\natexlab{}.
\newblock \showarticletitle{Measuring and testing dependence by correlation of
  distances}.
\newblock \bibinfo{journal}{\emph{The annals of statistics}}
  \bibinfo{volume}{35}, \bibinfo{number}{6} (\bibinfo{year}{2007}),
  \bibinfo{pages}{2769--2794}.
\newblock


\bibitem[\protect\citeauthoryear{Tang and Wang}{Tang and Wang}{2018}]%
        {tang2018personalized}
\bibfield{author}{\bibinfo{person}{Jiaxi Tang} {and} \bibinfo{person}{Ke
  Wang}.} \bibinfo{year}{2018}\natexlab{}.
\newblock \showarticletitle{Personalized top-n sequential recommendation via
  convolutional sequence embedding}. In \bibinfo{booktitle}{\emph{Proceedings
  of the Eleventh ACM International Conference on Web Search and Data Mining}}.
  \bibinfo{pages}{565--573}.
\newblock


\bibitem[\protect\citeauthoryear{Tian, Chang, Niu, Song, and Li}{Tian
  et~al\mbox{.}}{2022}]%
        {tian2022multi}
\bibfield{author}{\bibinfo{person}{Yu Tian}, \bibinfo{person}{Jianxin Chang},
  \bibinfo{person}{Yanan Niu}, \bibinfo{person}{Yang Song}, {and}
  \bibinfo{person}{Chenliang Li}.} \bibinfo{year}{2022}\natexlab{}.
\newblock \showarticletitle{When Multi-Level Meets Multi-Interest: A
  Multi-Grained Neural Model for Sequential Recommendation}. In
  \bibinfo{booktitle}{\emph{Proceedings of the 45th International ACM SIGIR
  Conference on Research and Development in Information Retrieval}}.
  \bibinfo{pages}{1632--1641}.
\newblock


\bibitem[\protect\citeauthoryear{Wang, Shen, Wang, Chen, Chen, and Wen}{Wang
  et~al\mbox{.}}{2022}]%
        {wang2022unbiased}
\bibfield{author}{\bibinfo{person}{Zhenlei Wang}, \bibinfo{person}{Shiqi Shen},
  \bibinfo{person}{Zhipeng Wang}, \bibinfo{person}{Bo Chen},
  \bibinfo{person}{Xu Chen}, {and} \bibinfo{person}{Ji-Rong Wen}.}
  \bibinfo{year}{2022}\natexlab{}.
\newblock \showarticletitle{Unbiased sequential recommendation with latent
  confounders}. In \bibinfo{booktitle}{\emph{Proceedings of the ACM Web
  Conference 2022}}. \bibinfo{pages}{2195--2204}.
\newblock


\bibitem[\protect\citeauthoryear{Wu, Wu, Qi, Liu, Tian, Li, He, Huang, and
  Xie}{Wu et~al\mbox{.}}{2022b}]%
        {wu2022feedrec}
\bibfield{author}{\bibinfo{person}{Chuhan Wu}, \bibinfo{person}{Fangzhao Wu},
  \bibinfo{person}{Tao Qi}, \bibinfo{person}{Qi Liu}, \bibinfo{person}{Xuan
  Tian}, \bibinfo{person}{Jie Li}, \bibinfo{person}{Wei He},
  \bibinfo{person}{Yongfeng Huang}, {and} \bibinfo{person}{Xing Xie}.}
  \bibinfo{year}{2022}\natexlab{b}.
\newblock \showarticletitle{Feedrec: News feed recommendation with various user
  feedbacks}. In \bibinfo{booktitle}{\emph{Proceedings of the ACM Web
  Conference 2022}}. \bibinfo{pages}{2088--2097}.
\newblock


\bibitem[\protect\citeauthoryear{Wu, He, Wang, Zhang, and Wang}{Wu
  et~al\mbox{.}}{2022a}]%
        {wu2022survey}
\bibfield{author}{\bibinfo{person}{Le Wu}, \bibinfo{person}{Xiangnan He},
  \bibinfo{person}{Xiang Wang}, \bibinfo{person}{Kun Zhang}, {and}
  \bibinfo{person}{Meng Wang}.} \bibinfo{year}{2022}\natexlab{a}.
\newblock \showarticletitle{A survey on accuracy-oriented neural
  recommendation: From collaborative filtering to information-rich
  recommendation}.
\newblock \bibinfo{journal}{\emph{IEEE Transactions on Knowledge and Data
  Engineering}} (\bibinfo{year}{2022}).
\newblock


\bibitem[\protect\citeauthoryear{Xia, Xu, Huang, Dai, and Bo}{Xia
  et~al\mbox{.}}{2021}]%
        {xia2021graph}
\bibfield{author}{\bibinfo{person}{Lianghao Xia}, \bibinfo{person}{Yong Xu},
  \bibinfo{person}{Chao Huang}, \bibinfo{person}{Peng Dai}, {and}
  \bibinfo{person}{Liefeng Bo}.} \bibinfo{year}{2021}\natexlab{}.
\newblock \showarticletitle{Graph meta network for multi-behavior
  recommendation}. In \bibinfo{booktitle}{\emph{Proceedings of the 44th
  International ACM SIGIR Conference on Research and Development in Information
  Retrieval}}. \bibinfo{pages}{757--766}.
\newblock


\bibitem[\protect\citeauthoryear{Xie, Ling, Wang, Wang, Xia, and Lin}{Xie
  et~al\mbox{.}}{2020}]%
        {DBLP:conf/ijcai/XieLWWXL20}
\bibfield{author}{\bibinfo{person}{Ruobing Xie}, \bibinfo{person}{Cheng Ling},
  \bibinfo{person}{Yalong Wang}, \bibinfo{person}{Rui Wang},
  \bibinfo{person}{Feng Xia}, {and} \bibinfo{person}{Leyu Lin}.}
  \bibinfo{year}{2020}\natexlab{}.
\newblock \showarticletitle{Deep Feedback Network for Recommendation}. In
  \bibinfo{booktitle}{\emph{Proceedings of the Twenty-Ninth International Joint
  Conference on Artificial Intelligence, {IJCAI} 2020}},
  \bibfield{editor}{\bibinfo{person}{Christian Bessiere}} (Ed.).
  \bibinfo{publisher}{ijcai.org}, \bibinfo{pages}{2519--2525}.
\newblock
\urldef\tempurl%
\url{https://doi.org/10.24963/ijcai.2020/349}
\showDOI{\tempurl}


\bibitem[\protect\citeauthoryear{Xue, Dai, Zhang, Huang, and Chen}{Xue
  et~al\mbox{.}}{2017}]%
        {xue2017deep}
\bibfield{author}{\bibinfo{person}{Hong-Jian Xue}, \bibinfo{person}{Xinyu Dai},
  \bibinfo{person}{Jianbing Zhang}, \bibinfo{person}{Shujian Huang}, {and}
  \bibinfo{person}{Jiajun Chen}.} \bibinfo{year}{2017}\natexlab{}.
\newblock \showarticletitle{Deep matrix factorization models for recommender
  systems.}. In \bibinfo{booktitle}{\emph{IJCAI}}, Vol.~\bibinfo{volume}{17}.
  Melbourne, Australia, \bibinfo{pages}{3203--3209}.
\newblock


\bibitem[\protect\citeauthoryear{Yu, Lin, Liu, Ge, Ou, and Qin}{Yu
  et~al\mbox{.}}{2021}]%
        {yu2021self}
\bibfield{author}{\bibinfo{person}{Wenhui Yu}, \bibinfo{person}{Xiao Lin},
  \bibinfo{person}{Jinfei Liu}, \bibinfo{person}{Junfeng Ge},
  \bibinfo{person}{Wenwu Ou}, {and} \bibinfo{person}{Zheng Qin}.}
  \bibinfo{year}{2021}\natexlab{}.
\newblock \showarticletitle{Self-propagation graph neural network for
  recommendation}.
\newblock \bibinfo{journal}{\emph{IEEE Transactions on Knowledge and Data
  Engineering}} (\bibinfo{year}{2021}).
\newblock


\bibitem[\protect\citeauthoryear{Yu, Lian, Mahmoody, Liu, and Xie}{Yu
  et~al\mbox{.}}{2019}]%
        {yu2019adaptive}
\bibfield{author}{\bibinfo{person}{Zeping Yu}, \bibinfo{person}{Jianxun Lian},
  \bibinfo{person}{Ahmad Mahmoody}, \bibinfo{person}{Gongshen Liu}, {and}
  \bibinfo{person}{Xing Xie}.} \bibinfo{year}{2019}\natexlab{}.
\newblock \showarticletitle{Adaptive User Modeling with Long and Short-Term
  Preferences for Personalized Recommendation.}. In
  \bibinfo{booktitle}{\emph{IJCAI}}. \bibinfo{pages}{4213--4219}.
\newblock


\bibitem[\protect\citeauthoryear{Zhang and Yang}{Zhang and Yang}{2018}]%
        {zhang2018overview}
\bibfield{author}{\bibinfo{person}{Yu Zhang} {and} \bibinfo{person}{Qiang
  Yang}.} \bibinfo{year}{2018}\natexlab{}.
\newblock \showarticletitle{An overview of multi-task learning}.
\newblock \bibinfo{journal}{\emph{National Science Review}}
  \bibinfo{volume}{5}, \bibinfo{number}{1} (\bibinfo{year}{2018}),
  \bibinfo{pages}{30--43}.
\newblock


\bibitem[\protect\citeauthoryear{Zheng, Gao, Chang, Niu, Song, Jin, and
  Li}{Zheng et~al\mbox{.}}{2022}]%
        {zheng2022disentangling}
\bibfield{author}{\bibinfo{person}{Yu Zheng}, \bibinfo{person}{Chen Gao},
  \bibinfo{person}{Jianxin Chang}, \bibinfo{person}{Yanan Niu},
  \bibinfo{person}{Yang Song}, \bibinfo{person}{Depeng Jin}, {and}
  \bibinfo{person}{Yong Li}.} \bibinfo{year}{2022}\natexlab{}.
\newblock \showarticletitle{Disentangling Long and Short-Term Interests for
  Recommendation}. In \bibinfo{booktitle}{\emph{Proceedings of the ACM Web
  Conference 2022}}. \bibinfo{pages}{2256--2267}.
\newblock


\bibitem[\protect\citeauthoryear{Zhou, Mou, Fan, Pi, Bian, Zhou, Zhu, and
  Gai}{Zhou et~al\mbox{.}}{2019}]%
        {zhou2019deep}
\bibfield{author}{\bibinfo{person}{Guorui Zhou}, \bibinfo{person}{Na Mou},
  \bibinfo{person}{Ying Fan}, \bibinfo{person}{Qi Pi}, \bibinfo{person}{Weijie
  Bian}, \bibinfo{person}{Chang Zhou}, \bibinfo{person}{Xiaoqiang Zhu}, {and}
  \bibinfo{person}{Kun Gai}.} \bibinfo{year}{2019}\natexlab{}.
\newblock \showarticletitle{Deep interest evolution network for click-through
  rate prediction}. In \bibinfo{booktitle}{\emph{Proceedings of the AAAI
  conference on artificial intelligence}}, Vol.~\bibinfo{volume}{33}.
  \bibinfo{pages}{5941--5948}.
\newblock


\bibitem[\protect\citeauthoryear{Zhou, Zhu, Song, Fan, Zhu, Ma, Yan, Jin, Li,
  and Gai}{Zhou et~al\mbox{.}}{2018}]%
        {zhou2018deep}
\bibfield{author}{\bibinfo{person}{Guorui Zhou}, \bibinfo{person}{Xiaoqiang
  Zhu}, \bibinfo{person}{Chenru Song}, \bibinfo{person}{Ying Fan},
  \bibinfo{person}{Han Zhu}, \bibinfo{person}{Xiao Ma},
  \bibinfo{person}{Yanghui Yan}, \bibinfo{person}{Junqi Jin},
  \bibinfo{person}{Han Li}, {and} \bibinfo{person}{Kun Gai}.}
  \bibinfo{year}{2018}\natexlab{}.
\newblock \showarticletitle{Deep interest network for click-through rate
  prediction}. In \bibinfo{booktitle}{\emph{Proceedings of the 24th ACM SIGKDD
  International Conference on Knowledge Discovery \& Data Mining}}.
  \bibinfo{pages}{1059--1068}.
\newblock


\bibitem[\protect\citeauthoryear{Zhu, Li, Liao, Wang, Guan, Liu, and Cai}{Zhu
  et~al\mbox{.}}{2017}]%
        {zhu2017next}
\bibfield{author}{\bibinfo{person}{Yu Zhu}, \bibinfo{person}{Hao Li},
  \bibinfo{person}{Yikang Liao}, \bibinfo{person}{Beidou Wang},
  \bibinfo{person}{Ziyu Guan}, \bibinfo{person}{Haifeng Liu}, {and}
  \bibinfo{person}{Deng Cai}.} \bibinfo{year}{2017}\natexlab{}.
\newblock \showarticletitle{What to Do Next: Modeling User Behaviors by
  Time-LSTM.}. In \bibinfo{booktitle}{\emph{IJCAI}}, Vol.~\bibinfo{volume}{17}.
  \bibinfo{pages}{3602--3608}.
\newblock


\end{thebibliography}
\nobalance

\end{document}